\documentclass[english]{article}

\usepackage[most]{tcolorbox}
\usepackage{graphicx} 
\usepackage{subfigure}
\usepackage{xcolor}
\usepackage{xspace}
\usepackage{booktabs}
\usepackage{url}
\usepackage{multirow}
\usepackage[linesnumbered, boxed]{algorithm2e}
\usepackage{amsmath, amsthm}
\usepackage{listings}

\usepackage[utf8]{inputenc} 
\usepackage[T1]{fontenc}    
\usepackage{hyperref}       
\usepackage{url}            
\usepackage{amsfonts}       
\usepackage{nicefrac}       
\usepackage{microtype}      

\usepackage{amssymb}
\usepackage{mylstlinebgrd}

\usepackage{geometry}
\usepackage[round]{natbib}
\usepackage{babel}
\usepackage{siunitx}
\usepackage{bm}
\usepackage{bbm}
\usepackage{caption}
\usepackage{appendix}

\definecolor{codegreen}{rgb}{0,0.6,0}
\definecolor{codegray}{rgb}{0.5,0.5,0.5}
\definecolor{codepurple}{rgb}{0.58,0,0.82}
\definecolor{backcolour}{rgb}{0.95,0.95,0.92}

\definecolor{eclipseBlue}{RGB}{42,0.0,255}
\definecolor{eclipseGreen}{RGB}{63,180,95}
\definecolor{eclipsePurple}{RGB}{175,0,25}
\definecolor{codewhite}{rgb}{0.70,0.70,0.70}

\definecolor{moderator}{HTML}{e3716e}
\definecolor{player1}{HTML}{bdb5e1}
\definecolor{player2}{HTML}{7ac7e2}
\definecolor{player3}{HTML}{f7df87}
\definecolor{player4}{HTML}{54beaa}
\definecolor{player5}{HTML}{2983b1}
\definecolor{player6}{HTML}{eca680}

\lstdefinestyle{mystyle}{
    backgroundcolor=\color{backcolour},   
    commentstyle=\color{codegreen},
    keywordstyle=\color{magenta},
    numberstyle=\tiny\color{codegray},
    stringstyle=\color{codepurple},
    basicstyle=\ttfamily\footnotesize,
    breakatwhitespace=false,         
    breaklines=true,
    breakindent=-3.5pt,
    captionpos=b,                    
    keepspaces=true,                 
    numbers=left,                    
    numbersep=5pt,                  
    showspaces=false,                
    showstringspaces=false,
    showtabs=false,                  
    tabsize=4,
    postbreak={
    	\mbox{
    		\lst@linebreakbgrd
    		\rotatebox[y=0.7ex]{180}{\color{black}$\Lsh\,$}
    	}
    },
}

\lstdefinelanguage{Dialog}{
	backgroundcolor=\color{backcolour},   
	keywordstyle=\color{magenta},
	numberstyle=\tiny\color{codegray},
	basicstyle=\ttfamily\footnotesize,
	breakatwhitespace=false,         
	breaklines=true,   
    breakindent=-3.5pt,
	captionpos=b,                    
	keepspaces=true,                 
	numbers=left,                    
	numbersep=5pt,                  
	showspaces=false,                
	showstringspaces=false,
	showtabs=false,                  
	tabsize=4,
	morecomment = [l][\color{eclipseGreen}\bfseries]{Assistant:},
    morecomment = [s][\color{eclipseGreen}\bfseries]{1.}{more.},
    morecomment = [l][\color{eclipseBlue}\bfseries]{User:},
}

\lstdefinelanguage{Werewolf}{
	backgroundcolor=\color{backcolour},   
	keywordstyle=\color{magenta},
	numberstyle=\tiny\color{codegray},
	basicstyle=\ttfamily\footnotesize,
	breakatwhitespace=false,         
	breaklines=true,   
    breakindent=-3.5pt,
	captionpos=b,                    
	keepspaces=true,                 
	numbers=left,                    
	numbersep=5pt,                  
	showspaces=false,                
	showstringspaces=false,
	showtabs=false,                  
	tabsize=4,
    morecomment = [l][\color{player1}\bfseries]{Player1:},
     morecomment = [l][\color{player2}\bfseries]{Player2 :},
     morecomment = [l][\color{player3}\bfseries]{Player3 :},
     morecomment = [l][\color{player4}\bfseries]{Player4 :},
     morecomment = [l][\color{player5}\bfseries]{Player5 :},
     morecomment = [l][\color{player6}\bfseries]{Player6 :},
     morecomment = [s][\color{moderator}\bfseries]{Moderator :}{\}},
}

\newcommand{\eg}{\emph{e.g.},\xspace}

\newcommand{\etc}{etc.\xspace}

\newcommand\figref[1]{Fig.~\ref{#1}}

\newcommand\lstref[1]{Listing~\ref{#1}}

\newcommand\secref[1]{Sec.~\ref{#1}}

\newcommand\appref[1]{Appendix~\ref{#1}}

\newcommand{\ours}{AgentScope\xspace}

\renewcommand{\lstlistingname}{Example}

\begin{document}

\title{AgentScope: A Flexible yet Robust Multi-Agent Platform}

\author{
Dawei Gao$^\dagger$, 
Zitao Li$^\dagger$,
Xuchen Pan$^*$,
Weirui Kuang$^*$,
Zhijian Ma$^*$,
Bingchen Qian$^*$,
\\
Fei Wei$^*$,
Wenhao Zhang$^*$,
Yuexiang Xie$^*$,
Daoyuan Chen$^*$, 
\\
Liuyi Yao,
Hongyi Peng,
Zeyu Zhang,
Lin Zhu,
Chen Cheng,
Hongzhu Shi,
\\
Yaliang Li$^\ddagger$, 
Bolin Ding$^\ddagger$, 
Jingren Zhou
\\
\\
\small{Alibaba Group}
}

\renewcommand{\thefootnote}{\fnsymbol{footnote}}
\footnotetext[2]{Co-first authors.} 
\footnotetext[1]{Equal contribution.}
\footnotetext[3]{Corresponding authors, email address: \{yaliang.li, bolin.ding\}@alibaba-inc.com}

\date{}

\maketitle

\begin{abstract}
With the rapid advancement of Large Language Models (LLMs), significant progress has been made in multi-agent applications.  
However, the complexities in coordinating agents' cooperation and LLMs' erratic performance pose notable challenges in developing robust and efficient multi-agent applications. 
To tackle these challenges, we propose \textbf{\ours}, a developer-centric multi-agent platform with message exchange as its core communication mechanism. 
The abundant syntactic tools, built-in agents and service functions, user-friendly  interfaces for application demonstration and utility monitor, zero-code programming workstation, and automatic prompt tuning mechanism significantly lower the barriers to both development and deployment. 
Towards robust and flexible multi-agent application, \ours provides both built-in and customizable fault tolerance mechanisms. 
At the same time, it is also armed with system-level support for managing and utilizing multi-modal data, tools, and external knowledge.
Additionally, we design an actor-based distribution framework, enabling easy conversion between local and distributed deployments and automatic parallel optimization without extra effort. 
With these features, \ours empowers developers to build applications that fully realize the potential of intelligent agents.
We have released \ours at \url{https://github.com/modelscope/agentscope}, and hope \ours invites wider participation and innovation in this fast-moving field. 
\end{abstract}

\section{Introduction}
Multi-agent systems, as upgraded extensions of single-agent systems, require collaborative efforts from multiple agents working in concert~\citep{agent-survey-1, agent-survey-2}. 
With the advancement of Large Language Models (LLMs)~\citep{instructGPT, gpt-4, llama, llama2}, multi-agent applications have made great progress in both research and industrial communities, including software engineering~\citep{meta-gpt}, society simulation~\citep{aitown}, and intelligent assistant~\citep{auto-gen, autogpt}.
Although significant progress has been made in multi-agent scenarios, there are still major challenges remaining in multi-agent application development.  

\textit{Developing a multi-agent application is more complex than creating a single-agent one.}
Unlike single-agent setups where an agent solely interacts with users, the development in the multi-agent scenario requires careful creation and management of multiple models and agents~\citep{agent-survey-1, agent-survey-2}, which poses high requirements for both versatility and handiness for a platform.
In particular, the following aspects feature the challenges: 1) Agents involved in a multi-agent application can specialize at different functions via different initial configurations; 2) A multi-agent application may require agents to be executed in a standardized operating procedure (SOP) or a more dynamic workflow; 3)The communication pattern between agents can be varying from one-to-one or broadcasting (e.g., a discussion group of agents).
As a result, developers expect a handy platform that can provide concise and clear programming patterns when taking care of all the aspects above, accelerating and facilitating the development cycle.
Achieving versatility and handiness simultaneously requires careful design and taking trade-offs, and it remains a persistent goal for all multi-agent platform designs.

\textit{Aberrations are tinderboxs in a multi-agent system.}
Although LLMs have advanced rapidly, they still struggle with issues like hallucination~\citep{hall1, hall2} and inadequate instruction-following~\citep{follow1, follow2}.
Besides, an agent can be equipped with various tools, but those tools introduce additional uncertainties (e.g., accessibility to a database or the search engine).
From the perspective of multi-agent system robustness, any unexpected error or response can propagate to the whole system, causing a series of cascading effects if not handled properly.
Thus, it is crucial for multi-agent applications to autonomously detect and handle unexpected responses from LLMs. 
While LLMs may assist in identifying and managing these errors, it remains a challenge to determine whether they can resolve errors on their own and to automatically provide the necessary information for error correction. 
Consequently, designing fault-tolerant that incorporate LLMs is a key challenge in the development of multi-agent applications. 

\textit{Supporting agents with multi-modal data, tools, and external knowledge is highly systematic.} 
Besides generating answers with LLMs, agents are expected to be more versatile, including generating and handling multi-modal data~\citep{multimodal, dalle3}, preparing and invoking functions as tools~\citep{react, shen2024small}, managing external knowledge banks, and using the retrieved knowledge for augmentation generation~\citep{lewis2020retrieval}.
However, integrating these functionalities in multi-agent applications requires a comprehensive and systematic approach. 
Supporting multi-modal content is a complex endeavor, necessitating considerations for data storage, presentation, user interaction, message transmission, and communication. 
Tool utilization of agents requires unifying the function calling pattern and output parsing, prompting to instruct LLMs, and designing reasoning mechanisms to ensure the tasks can be accomplished step by step.
As for external knowledge, beyond the retrieval-augmented generation (RAG) techniques, we need to consider how to efficiently share and manage the knowledge in multi-agent scenarios while leaving enough flexibility for retrieval strategies.
While some existing works investigate how those techniques individually work within specialized agent systems, general platform-level programming interfaces remain absent.

\textit{Distributed applications bring extra programming difficulties and system design challenges.}
An industrial-oriented scenario for multi-agent applications is that the agents are owned by different organizations and run on different machines because the agents are equipped with unique private knowledge or patented tools.
Developing such applications usually requires the developers to have professional knowledge of distributed system programming and optimization in the design phase.
Besides, distributed applications usually require a great extra effort in the development and testing, especially when debugging and diagnosing issues spread across distributed processes or agents.
Moreover, integrating advanced features like multi-modal data processing poses additional challenges in a distributed setting, when the agents require different time to accomplish the sub-tasks or the generated contents are very heterogeneous. 
Poor distributed system design can result in excessive communication overhead between agents. 
Therefore, building distributed multi-agent applications requires the large efforts of experienced developers and a high barrier for beginners to migrate their prototypes to a distributed style for optimal efficiency.

To tackle the aforementioned challenges, we introduce \textbf{\ours}, a novel multi-agent platform designed for developers with varying levels of expertise. 
\ours is well-designed with a message exchange communication mechanism that embodies great usability, robustness, and efficiency.
We underscore the salient features of \ours as follows:

\textbf{Exceptional Usability for Developers}.
\ours is designed with a fundamental emphasis on ease of use, particularly for developers with varying levels of expertise. 
By implementing a procedure-oriented message exchange mechanism, \ours ensures a smooth learning curve on multi-agent application development. 
To alleviate the programming burdens, \ours offers an extensive suite of syntactic utilities, including various pipelines and an information-sharing mechanism.
Besides programming with our framework, we also improve usability by providing a \emph{zero-code drag-and-drop programming workstation}, which can enable those with limited Python programming experience to build their own applications with little effort.
Compared with building the skeleton of the application, prompt tuning can be a more time-consuming stage in multi-agent application development.
In \ours, we equip our agents with a set of \emph{automatic prompt tuning} mechanisms to relieve such burden.
Coupled with rich built-in resources and integrated user interaction modules, \ours makes building a multi-agent application much more enjoyable than ever. 

\textbf{Robust Fault Tolerance for Diverse LLMs and APIs}.
As the scale and scope of models and APIs expand, a robust fault-tolerance mechanism in multi-agent applications becomes paramount. 
\ours integrates a comprehensive service-level retry mechanism to maintain API reliability. 
\ours is equipped with a set of rule-based correction tools to handle some obvious formatting problems in the responses of LLMs. 
Moreover, \ours offers customizable fault tolerance configurations, enabling developers to tailor their own fault tolerance mechanism through parameters like \texttt{parse\_func}, \texttt{fault\_handler}, and \texttt{max\_retries}.
While admittedly, not all the errors can be handled by the aforementioned mechanism, we propose a logging system with customized features for multi-agent applications as the last safeguard for \ours. 

\textbf{Extensive Compatibility for Multi-Modal, Tools, and External Knowledge}. 
With the remarkable progress of large-scale multi-modal models, \ours supports multi-modal data (\eg texts, images, audio, and videos) in dialog conversation, message transmission, and data storage. 
Specifically, \ours decouples multi-modal data transmission from storage and employs \emph{a lazy loading strategy} by providing a unified URL-based attribute in messages. 
During message transmission, \ours only attaches a URL to the message, and the multi-modal data is loaded only when necessary, such as when being rendered in web UI or invoked by model wrappers. 
For tool usage, \ours provides a component, called \emph{service toolkit}, as a one-step solution for tool usage, including function preprocessing, prompt engineering, reasoning, and response parsing with fault-tolerance features.
To support efficient external knowledge usage, \ours provides end-to-end, highly configurable, and sharable knowledge processing modules for retrieval-augmented generation (RAG), from data preprocessing to customizable retrieval.

\textbf{Optimized Efficiency for Distributed Multi-Agent Operations}.
Recognizing the vital importance of distributed deployment, \ours introduces an actor-based distributed mechanism that enables centralized programming of complex distributed workflows, and automatic parallel optimization. 
Particularly, the workflows for local and distributed deployments is a exactly the same one, indicating negligible overhead when migrating applications between centralized and distributed environments. 
With such a distribution framework, \ours empowers developers to concentrate on the application design rather than implementation details. 

\noindent
\paragraph{Summary}
To summarize, \ours, a novel multi-agent platform proposed for flexibility and robustness, includes the following advanced features:
\begin{enumerate}
    \item 
    \ours provides a procedure-oriented message exchange mechanism with a set of syntactic features to facilitate multi-agent programming, a zero-code drag-and-drop programming workstation, and a set of automatic prompt tuning mechanisms.
    
    \item 
    The fault tolerance designs of \ours enable developers to handle errors elegantly for their applications.
    
    \item 
    The support for the multi-modal applications reduces the overheads of heterogeneous data generation and transmission. The service toolkit component facilitates the tool usage of agents in \ours, and the knowledge processing modules provide a flexible solution for agents to handle different information.
    
    \item
    The actor-based distributed mode of \ours can help develop efficient and reliable distributed multi-agent applications seamlessly.
\end{enumerate}

\noindent\paragraph{Roadmap}
In the following sections, we navigate through the core components and capabilities of \ours, showcasing its role in advancing the development and deployment of multi-agent applications. Section \ref{sec:overview} provides an overview, while Section \ref{sec:usability} focuses on the user experience.
Section \ref{sec:tolerance} introduces the fault tolerance mechanism in \ours.
Sections \ref{sec:mutil-modal}, \ref{sec:tools}, and \ref{sec:rag} cover the multi-modal support, tool usage, and retrieval-augmented generation modules in \ours. 
Section \ref{sec:distribute} presents our platform's support for distributed multi-agent applications.
Use cases are presented in Section \ref{sec:app}, related work is summarized in Section \ref{sec:related_work}, and concluding thoughts are recorded in Section \ref{sec:conclusion}.

\section{Overview}
\label{sec:overview}

\subsection{Basic Concepts in \ours}
\label{sec:concept}

This section introduces the primary concepts in \ours: {\it message}, {\it agent}, {\it service}, and {\it workflow}.
These four concepts are throughout the platform and all multi-agent applications based on it. 

\begin{itemize}
\item \textbf{Message}:
Messages serve as the carriers for information exchange in multi-agent conversations, encapsulating the source and content of the information. In \ours, messages are implemented as Python dictionaries with two mandatory fields (\textit{name} and \textit{content}) and an optional field (\textit{url}). 
The \textit{name} field records the name of the agent that generates the message, and the \textit{content} field contains the text-based information generated by the agent.
The \textit{url} field is designed to hold the Uniform Resource Locator (URL), which typically links to multi-modal data, such as images or videos.
Messages with this field are particularly relevant for interactions with agents that can process and generate multi-modal content. 
Each message is uniquely identified by an auto-generated UUID and timestamp, ensuring traceability. 
\lstlistingname~\ref{lst:msg} shows how the messages can be created, 
serving as atoms in the inter-agent communication of \ours.

\begin{lstlisting}[language=Python, caption={Illustrative examples of message creation in \ours.}, label={lst:msg}, float=h]
from agentscope.message import Msg

msg1 = Msg("Alice", "Hello!")
msg2 = Msg(
    name="Bob",
    content="How do you find this picture I captured yesterday?",
    url="https://xxx.png"
)
\end{lstlisting}

\item \textbf{Agent}: 
Agents are the primary actors within multi-agent applications, acting as conversational participants and executors of tasks. In \ours, agent behaviors are abstracted through two interfaces: the \textit{reply} and \textit{observe} functions. The \textit{reply} function takes a message as input and produces a response, while the \textit{observe} function processes incoming messages without generating a direct reply. 
The interplay between agents and messages, as shown in \lstlistingname~ \ref{lst:agent}, forms the operational basis of \ours and is essential for developers to model complex interactions in multi-agent LLMs.

\begin{lstlisting}[language=Python, caption={Demonstration of message exchange between agents in \ours.}, label={lst:agent}, float=h]
# agent1 and agent2 are two initialized agents, for example
# agent1, agent2 = DialogAgent(...), DialogAgent(...)
msg1 = agent1()
msg2 = agent2(msg1)
\end{lstlisting}


\item \textbf{Workflow}:
Workflows represent ordered sequences of agent executions and message exchanges between agents, analogous to computational graphs in TensorFlow, but with the flexibility to accommodate non-DAG structures. 
Workflows define the flow of information and task processing among agents, facilitating parallel execution and efficiency improvements. 
This concept is essential for designing multi-agent systems that interact with LLMs, as it allows for the coordination of complex, interdependent tasks.

\item \textbf{Service Functions and Tools}: 
Note that service functions are closely related to but \emph{different} from the concept, tools, in the context of agent design in \ours.
Service functions refer to the functional APIs that return a formatted output \texttt{ServiceResponse}, while tools refer to processed services functions with functionality descriptions and necessary input parameters prepared.
We introduce these two concepts in \ours because LLMs require help to invoke service functions as tools.
One observation is that LLMs may need help understanding the functionalities of the service functions precisely and demand more descriptive information to make accurate decisions.
Meanwhile, LLMs can not (reliably) fill in some input parameters of the APIs, such as the API keys of Bing and Google Search.
As a result, \ours defines tools as processed service functions.
\end{itemize}

\begin{figure}
    \centering
    \includegraphics[width=0.6\linewidth]{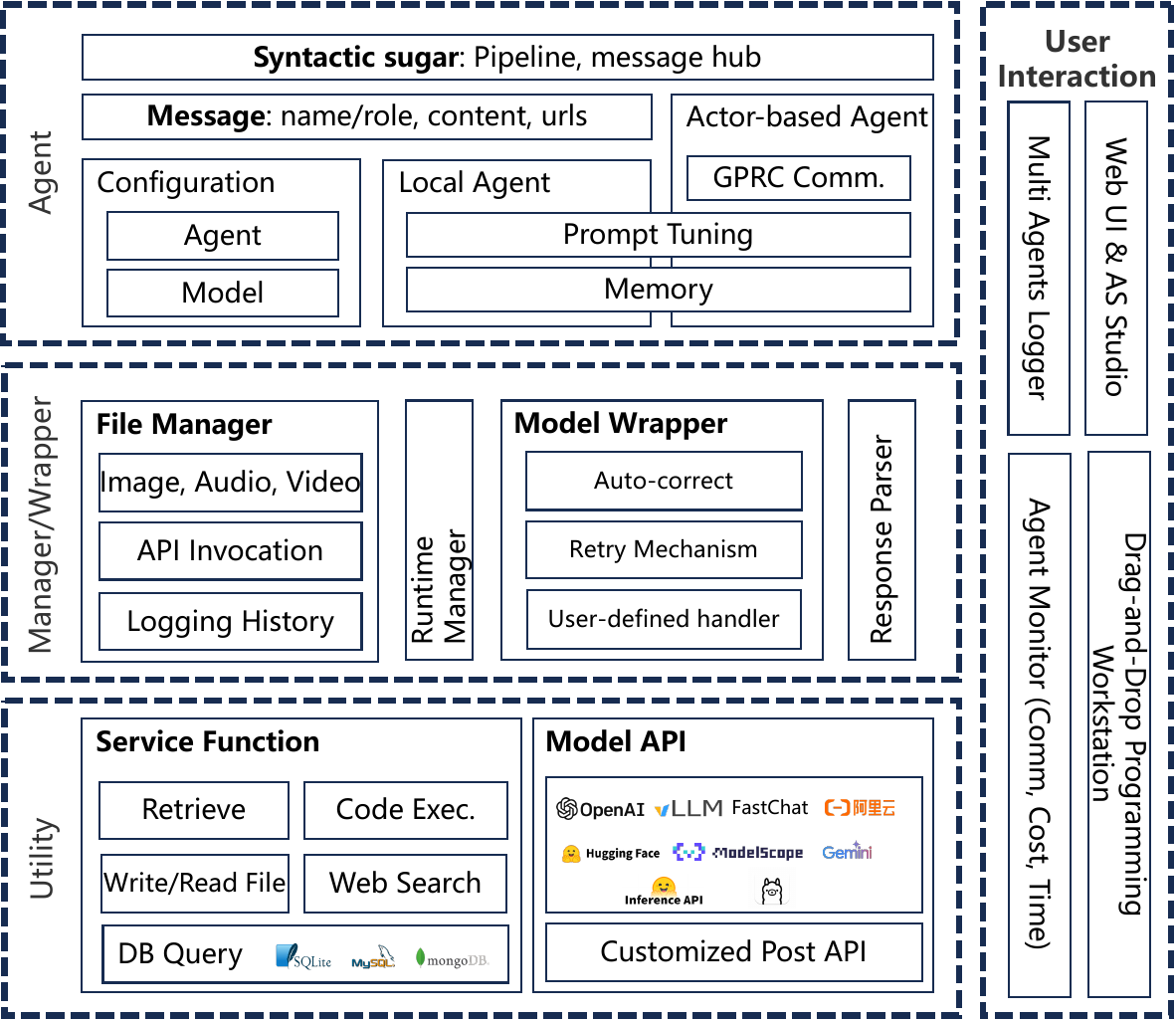}
    \caption{Architecture of \ours.}
    \label{fig:arch}
\end{figure}

\subsection{Architecture of \ours}
We present \ours as an infrastructural platform to facilitate the creation, management, and deployment of multi-agent applications integrated with LLMs. 
The architecture of \ours comprises three hierarchical layers and a set of user interaction interfaces, as shown in \figref{fig:arch}.
These layers provide support for multi-agent applications from different levels, including elementary and advanced functionalities of a single agent (utility layer), resources and runtime management (manager and wrapper layer), and agent-level to workflow-level programming interfaces (agent layer).
\ours introduces intuitive abstractions designed to fulfill the diverse functionalities inherent to each layer and simplify the complicated inter-layer dependencies when building multi-agent systems.
Furthermore, we offer programming interfaces and default mechanisms to strengthen the resilience of multi-agent systems against faults within different layers.

\begin{itemize}
\item \textbf{Utility Layer}: As the platform's foundation, the utility layer in \ours provides essential services to support the core functionalities of agents. This layer abstracts the complexity of underlying operations, such as model API invocation and service functions including code execution and database operations, allowing agents to focus on their primary tasks. 
\ours's utility layer is designed with ease of use and robustness as its utmost priority, supporting versatile operations in multi-agent systems and providing \emph{built-in} autonomous retry mechanisms for exception and error handling against unexpected interruptions.

\item \textbf{Manager and Wrapper Layer}: As an intermediary, the manager and wrapper abstraction layer manages the resources and API services, ensuring high availability of resources and providing resistance to undesired responses from LLMs. 
Unlike the utility layer, which provides default handlers, the manager and wrapper layer also offers customizable interfaces for fault tolerance controls depending on developers' needs and the specific requirements of the application. 
This layer is responsible for maintaining the operational integrity of the agents, a crucial aspect for LLMs to perform consistently under diverse conditions. Detailed elaboration on the fault tolerance mechanisms is provided in Section \ref{sec:tolerance}.

\item \textbf{Agent Layer}: At the core of \ours lies the agent abstraction, which forms the backbone of the multi-agent workflow and is the primary entity responsible for interaction and communication. This layer is designed to facilitate the construction of intricate workflows and enhance usability, reducing the programming burden on developers. 
By integrating streamlined syntax and tools, \ours empowers developers to concentrate on the implementation and optimization of agent-based applications that leverage the capabilities of LLMs. 
The programming features and syntactic sugars are introduced in Section \ref{sec:usability} with more details.

\item \textbf{User interaction}: In addition to the layered architecture, \ours provides multi-agent oriented interfaces such as an annotated terminal presenting basic information, Web UI monitoring the system, a Gradio-base~\citep{abid2019gradio} interface that can change a command line application to a graphical one with only one step and a drag-and-drop zero-code programming workstation (Figure~\ref{fig:workstation}). These interfaces allow developers to effortlessly monitor the status and metrics of the application, including agent communication, execution timing, and financial costs.

\end{itemize}

Collectively, the layered constructs of \ours provide the essential building blocks for developers to craft bespoke multi-agent applications that leverage the advanced capabilities of large language models. The subsequent section will delve into the features of \ours that enhance the programming experience for multi-agent application development.

\section{High Usability}
\label{sec:usability}

The design of \ours prioritizes usability, aiming to streamline the development process for multi-agent with LLMs and to ensure a smooth interaction experience for both users and developers. 
This section delves into how \ours flattens the learning curve and enhances the programmer's experience by introducing intuitive concepts and features that facilitate the creation of complex multi-agent applications.

\subsection{Syntactic Sugar for Multi-Agent Workflows}
Leveraging basic concepts introduced in Section \ref{sec:concept}, developers are empowered to construct sophisticated multi-agent applications. Nonetheless, directly coding each agent's message exchange can become cumbersome, as shown in \lstlistingname~\ref{lst:without_pipeline}. Recognizing this, \ours introduces two syntactic utilities: pipelines and message hubs, to abstract away the complexity and minimize repetition.

\begin{lstlisting}[language=Python, float=h, caption={Example of programming a sequential workflow with basic concepts in \ours.}, label={lst:without_pipeline}]
# set up agents: agent1 to agent5
# ...

msg = agent1(Msg ( " Alice " , " Hello ! " ))
msg = agent2(msg)
msg = agent3(msg)
msg = agent4(msg)
msg = agent5(msg)
\end{lstlisting}

\paragraph{Pipeline Abstraction} 
The pipeline abstraction reduces repetitive coding by encapsulating patterns of message transmission, including {\it sequential}, {\it conditional}, and {\it iterative} exchanges, into simple and reusable components. With these pipelines, developers can focus on the logic of agent interactions rather than the boilerplate code. \lstlistingname~\ref{lst:pipeline} illustrates how pipelines can be employed in both functional and object-oriented styles to create a clear and concise agent workflow.
Besides the sequential pipeline in the example, \ours also provides if-else, switch, while-loop, and for-loop pipelines, facilitating the programming of the multi-agent interactions.

\begin{lstlisting}[language=Python, float=h, caption={Using functional and object sequential pipeline to construct workflow in \ours.}, label={lst:pipeline}]
# set up agents: agent1 to agent5
# ...
from agentscope.pipelines import SequentialPipeline
from agentscope.pipelines.functional import sequentialpipeline

# using functional pipeline
x = sequentialpipeline([agent1, agent2, agent3, agent4, agent5], x)

# using object pipeline
pipe = SequentialPipeline([agent1, agent2, agent3, agent4, agent5])
x = pipe(x)
\end{lstlisting}  

\paragraph{Message Hub for Agent Communication}
In multi-agent systems, especially when integrated with LLMs, efficiently managing communication among a group of agents is essential. The message hub in \ours serves as a broadcast mechanism that simplifies group interactions. Developers can initiate a message hub by defining participating agents and can include initial broadcast messages. 
When new messages are generated by the agents within the message hub, they are automatically disseminated to other participants, as demonstrated in \lstlistingname~\ref{lst:msghub}. 
This abstraction is particularly useful for multi-agent scenarios involving LLMs, where dynamic and contextually rich conversations are commonly observed~\citep {du2023improving}.

\begin{lstlisting}[language=Python, float=h, caption={Using message hub with \ours.}, label={lst:msghub}]
# set up agents: agent1 to agent4
# ...

greeting = Msg("host", "Welcome to the message hub!")

with msghub(participant=[agent1, agent2, agent3], 
            announcement=greeting) as hub:
    # Message will be broadcast to agent2 and agent3 automatically
    agent1()
    
    # Delete agent2 from the message hub
    hub.delete(agent2)
    
    # Add agent4 into the message hub
    hub.add(agent4)
    
    # Broadcast message
    hub.broadcast(Msg("host", "Welcome agent4 to join the hub!"))
\end{lstlisting}  

\subsection{Resource-Rich Environment for Agent Development}
To further enhance usability, \ours is equipped with a rich set of built-in resources, including services, dedicated agents, and pre-configured examples. These resources are designed to reduce the initial setup effort and enable rapid prototyping and deployment of multi-agent LLM systems.

\paragraph{Comprehensive Service Integration}
\ours integrates various service functions, such as web search, database querying, and code execution, to support the tool usage capabilities of agents. 
These service functions are essential for building helpful agents with LLMs, as agents often need to draw information from external sources or execute tasks that go beyond the equipped LLMs' internal knowledge. \lstlistingname~\ref{lst:web_search} showcases the seamless conversion of a service into an OpenAI-Compatible JSON format, simplifying the integration process for developers.

\begin{lstlisting}[language=Python, float=t, caption={ Converting web search service into the function and JSON format dictionary that agent can use.}, label={lst:web_search}]
from agentscope.service import ServiceFactory, web_search

bing_search, func_json = ServiceFactory.get(web_search, engine="bing", api_key="xxx", num_results=10)

print(func_json)
# {
#     "name": "web_search",
#     "description": "Searching the given question with bing.",
#     "parameters": {
#         "type": "object",
#         "properties": {
#             "type": "object",
#             "properties": {
#                 "question": {
#                     "type": "string",
#                     "description": "The string question to search in Bing."
#                 }
#             }
#         }
#     }
# }

searching_result = bing_search("What's the date today?")
\end{lstlisting}

\paragraph{Pre-built Agent Templates}
As cataloged in Table \ref{tab:agents}, \ours offers pre-built agents and ready-to-use components for tasks like dialogue management, user proxying, multi-modal data handling, and distributed deployment. These templates serve as starting points for developers to customize and extend, significantly accelerating the development of multi-agent LLM applications.

\begin{table}[ht]
    \centering
    \begin{tabular}{l|l}
        \toprule
        Agent Name                 &    Function            \\ \hline
        UserAgent                  &    The proxy of the user.  \\
        DialogAgent                &    A general dialog agent, whose role can be set by system prompt.\\
        DictDialogAgent            &    A dictionary version dialog agent, who responds in Python dictionary format.  \\
        ReActAgent   & An agent that can reason and use tools \\
        ProgrammerAgent     &    An agent that can write and execute Python code.   \\
        TextToImageAgent       &    An agent that generates images according to the requirements.  \\
        RpcUserAgent               &    A distributed version user proxy.   \\
        RpcDialogAgent             &    A distributed version DialogAgent.   \\
        \bottomrule
    \end{tabular}
    \caption{Some examples of built-in agents and their functions in \ours.}
    \label{tab:agents}
\end{table}

\subsection{Multi-Agent Oriented Demonstration Interfaces}
\label{sec:interactive}

\begin{figure}[ht]
    \centering
    \includegraphics[width=0.75\linewidth]{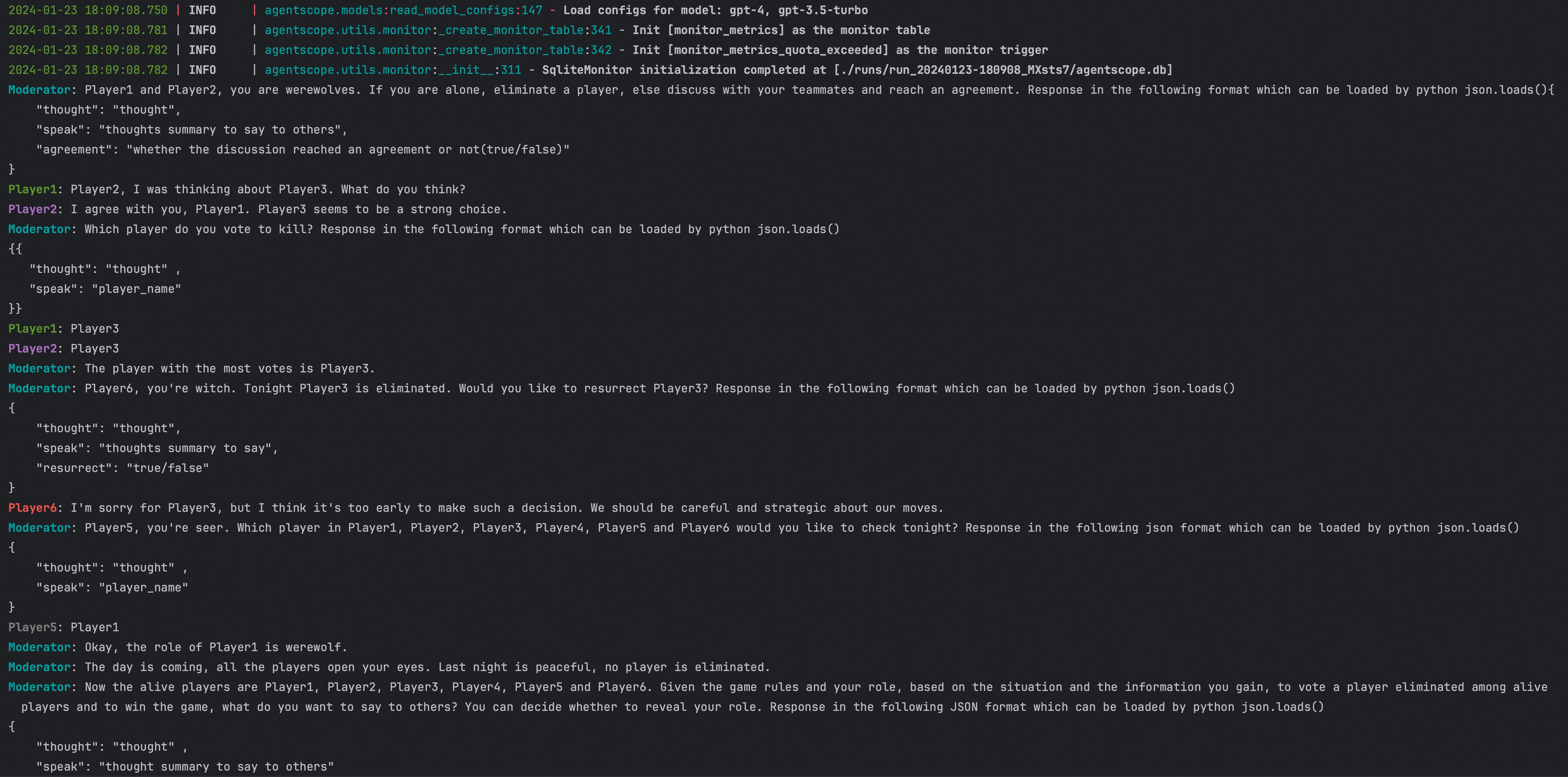}
    \caption{The dialogue history of a werewolf game in \ours.}
    \label{fig:modal_terminal}
\end{figure}

\begin{figure}[ht]
    \centering
    \fbox{\includegraphics[width=0.86\linewidth]{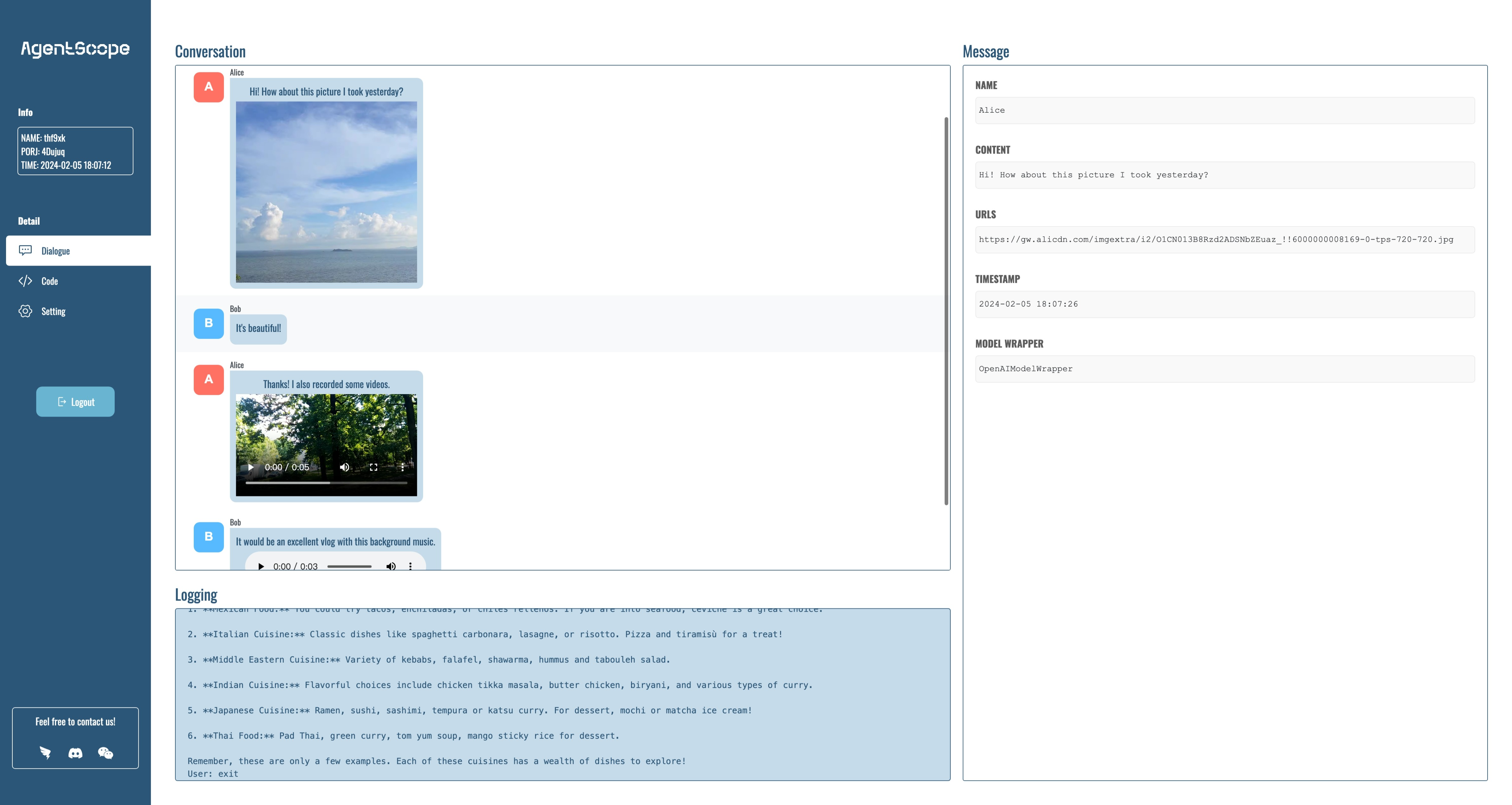}}
    \caption{Multi-modal interactions between agents in web UI.}
    \label{fig:modal_webui}
\end{figure}

Furthermore, \ours introduces interaction interfaces tailored for multi-agent systems, as illustrated in Figures \ref{fig:modal_terminal} and \ref{fig:modal_webui}. These interfaces provide a rich multi-modal experience, crucial for systems incorporating LLMs that handle diverse data types.

\paragraph{Agent Differentiation in User Interfaces}
To facilitate user interaction with multiple agents, \ours assigns unique colors and icons to each agent, enhancing clarity and visual distinction in both terminal and web UI (\figref{fig:modal_webui}). The ``first-person perspective'' feature allows users to experience interactions from the viewpoint of a specified agent, aligning with their role in the application, such as in a game scenario. This feature not only enriches the multi-agent experience but also mirrors the nuanced interactions that occur in human-agent and agent-agent dialogues within LLM systems.

\paragraph{Monitoring and Cost Management}
A vital aspect of deploying LLMs at scale is resource management. \ours includes a monitoring module that tracks model and API usage, as well as calculating financial costs. Developers can customize metrics and set budget limits, receiving automatic alerts when thresholds are approached or exceeded. This proactive cost management is particularly important for LLMs that may incur high computational expenses.

\paragraph{AgentScope Gradio Interface}
Once you have a multi-agent application, executing it in the terminal may be a concise choice but lacks attraction.
In \ours, we provide a powerful Gradio-based interface that is compatible with all \ours applications as long as there is a \texttt{main} function as the application's entry point. 
For example, if the main function of the application is in \texttt{application.py} file, then running ``$\texttt{as\_studio application.py}$'' can build a Gradio application with a graphical user interface and support multi-modal content upload and presentation.

\subsection{Towards Graphical Application Development}

\begin{figure}[ht]
    \centering
    \fbox{\includegraphics[width=0.86\linewidth]{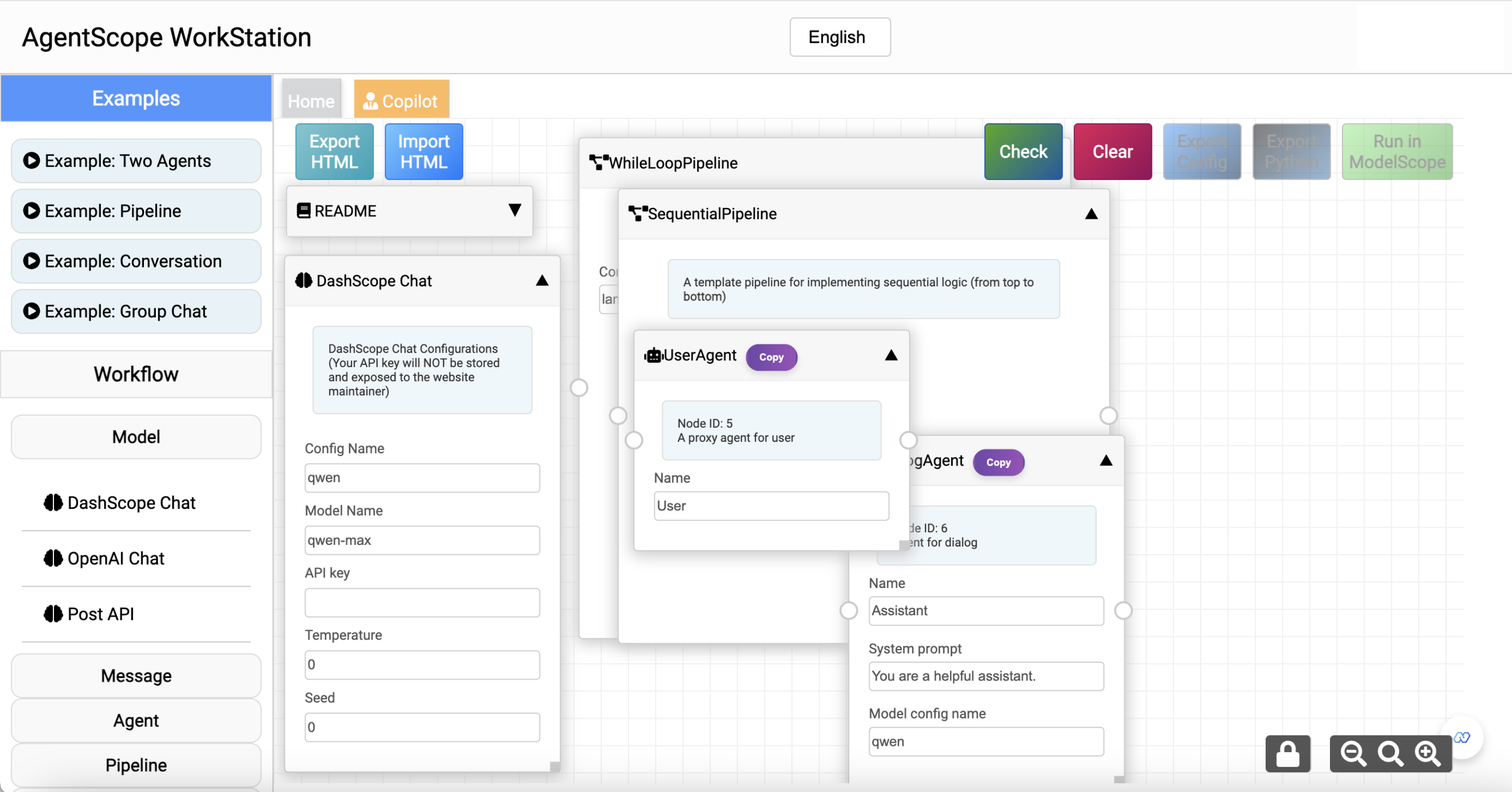}}
    \caption{Drag-and-drop programming \href{https://agentscope.aliyun.com/}{workstation}.}
    \label{fig:workstation}
\end{figure}

The design mentioned above provides massive convenience for those familiar with Python programming to quickly develop their multi-agent applications. However, \ours takes a step further. \ours provides a drag-and-drop online workstation on which developers only need to drag the module blocks to compose an application; then, the workstation can generate a configuration file of the application in JSON or even a piece of Python code. With this feature, those with limited experience with Python programming can build their multi-agent application without writing any Python code, while those familiar with Python can instantly obtain a piece of draft code ready for further customization.
A screenshot of the online workstation is shown in \figref{fig:workstation}, and the idea supporting this implementation is illustrated as follows.

\paragraph{Expressing Multi-agent application with nodes in directed acyclic graph (DAG).}
Based on the highly modular design of our basic infrastructure, all the key components can be represented as a node, and an application can be built by constructing a directed cycle graph (DAG). The execution of the application is equivalent to triggering and running the nodes in the graph following the traversing order of DAG. Following the traditional terms, we name such DAG execution as a \emph{workflow} and name the nodes in the workflow as \emph{workflow nodes}.  
According to their functionality, the workflow nodes are categorized into six different types: model nodes, agent nodes, pipeline nodes, service nodes and copy nodes.  
\begin{itemize}
    \item 
    {\bf Model nodes}: Model nodes are designed to be relatively independent of the DAG. They correspond to the model configurations in \ours and work as entries to let users configure their models (LLMs, embedding models, or multi-modal models) and maintain such information for all the nodes in the following workflow that need to use the model.
    
    \item 
    {\bf Service (tool) nodes}: These nodes correspond to the services available in \ours. Some of them require additional information to set up, such as Google search and Bing search, which require API keys; others can be used directly. 
    
    \item 
    {\bf Agent nodes}: As the name suggests, agent nodes represent the agents in \ours, which means users need to decide the models, agent name and system prompts for the agent. 

    \item 
    {\bf Pipeline nodes}: The pipeline node includes the operators of \ours, including the message hub and the pipelines (sequential, for-loop, while-loop, etc.). With such nodes, DAG representations can be as concise as Python programming.
    
    \item 
    {\bf Message node}: The message node is designed for cases where some initial messages are needed, such as the announcement (initial message) for the message hub.
    
    \item 
    {\bf Copy node}: The copy node is a special kinds of node that replicate the results of a parent node when its output is needed for multiple subsequent operations.
\end{itemize}

\paragraph{Execute DAG with JSON or compile to Python.}
With the nodes above, developers can build applications by composing DAGs. However, the DAG is highly UI-dependent. Although a DAG can be represented in some formats (e.g., JSON format recording each node's information and execution dependency), we still need to ensure it is as reusable as other applications. To overcome this, \ours is equipped with a data structure called ASDiGraph, which provides two solutions based on it. 
\begin{itemize}
    \item 
    {\bf Direct-run}: Given a JSON file recording the DAG information, ASDiGraph can parse DAG information and sort the nodes in topological order. With these sorted nodes, the run function of ASDiGraph can execute them in order and feed the predecessor's output to their successors as an application is executed step by step.

    \item 
    {\bf To-Python compiler}: The second solution is to translate the JSON file to a Python script. With the highly modularized components of \ours, the key idea is to rely on internal mappings of the functionality, required inputs, and expected outputs to small pieces of Python code. Specifically, each node contains Python code for importing dependent modules, initiating models or agents, and executing the application logic. ASDiGraph first groups the pieces of importing code and initiating code, and then it composes the pieces of execution code following the topological order.
    Therefore, users will obtain a complete Python script after the ASDiGraph finishes compilation.
\end{itemize}

\subsection{Automatic Prompt Tuning}
\label{sebsec:prompt_tuning}
For a multi-agent system that utilizes LLMs for generation, writing an appropriate prompt requires significant human effort and expertise~\citep{pryzant2023automatic}, which motivates us to provide automatic prompt generation and tuning in \ours for its high usability. Specifically, \ours allows users to generate prompts based on a simple description of the agent in natural language, update prompts according to contexts, and enable in-context learning.

\paragraph{System Prompt Tuning} When an agent is created, a system prompt should be associated with the agent to define its roles and responsibilities for following human instructions. For example, a Programmer Agent might be prompted as ``You are proficient in writing and executing Python code''. Meanwhile, a detailed and informative system prompt can improve agent performance and ensure that the agent performs as expected, such as ``You are proficient in writing and executing Python code. You prefer to write the code in a modular fashion and provide unit tests for each module''. With \ours, users only need to provide a simple description of the agent when creating the agents, and \ours can automatically generate such helpful system prompts using built-in tools based on LLMs, as shown in Example~\ref{lst:prompt_generation}.
\begin{lstlisting}[language=Python, float=h, caption={Initialize a programmer agent with automatic system prompt generation.}, label={lst:prompt_generation}]
# set up agents with automatic prompt generation
# ...
from agentscope.agents import ProgrammerAgent

# Load model configs
agentscope.init(model_configs="model_configs.json")

# Create a programmer agent 
programmer_agent = ProgrammerAgent(name="assistant", auto_sys_prompt=True,
                        model_config_name="my_config",
                        sys_prompt="an assistant that can write Python code")
\end{lstlisting}

Besides, \ours provides interfaces for system prompt updates, which include manually setting by users or automatically adjusting based on the context. As a promising future direction, meta-prompting techniques~\citep{pryzant2023automatic, suzgun2024meta} can also be integrated into \ours, which might involve integrating an evaluator to provide guidance for automatic prompt optimization.

\paragraph{In-Context Learning} Providing multiple demonstrations to the LLMs can greatly enhance their ability to follow instructions, particularly when we want them to complete specific downstream tasks~\citep{dai2023gpt, wei2022emergent}. \ours provides a simple switch to turn on/off the in-context learning behavior for agents that utilize LLMs. When users choose to apply in-context learning, they only need to provide demonstration candidates and configure how to match the most suitable ones, as illustrated in Example~\ref{lst:icl}. \ours offers several widely-used and useful matching approaches, such as random selection, similar questions, and similar answers, and allows for user customization.

\begin{lstlisting}[language=Python, float=h, caption={Enable in-context learning when creating an agent.}, label={lst:icl}]
# set up agents with in-context learning
# ...
from agentscope.agents import ReActAgent
from agentscope.utils.common import load_demo_data

# Load model configs
agentscope.init(model_configs="model_configs.json")

# Load demonstrations
react_pairs = load_demo_data("my_demos.txt")

# Create a reAct agent 
react_agent = ReActAgent(name="react_agent", enable_icl=True, 
                         demos=react_pairs, matching_approach="random")
\end{lstlisting}
\section{Fault-Tolerant Mechanisms}
\label{sec:tolerance}

In the realm of multi-agent systems, particularly those interfacing with diverse open-source LLMs with various instruction-following capabilities, fault tolerance is a key property to ensure seamless operation.
\ours is engineered to autonomously handle a wide range of errors with minimal human intervention required, drawing upon a comprehensive fault-tolerant infrastructure that is acutely aware of the complexities involved in multi-agent coordination and LLM dependencies.

\paragraph{Error Classification and Handling Strategies}
Our approach begins with a methodical classification of errors into distinct levels, each with tailored handling strategies:

\begin{itemize}
    \item 
    \textbf{Accessibility errors}:
    In \ours, an agent's functionalities rely on different kinds of services, but those services may be subject to temporary inaccessible errors.
    These errors may be caused by model instability or network conditions.
    For example, the model APIs may return a timeout error when there is traffic congestion during busy hours, or a database on a remote machine may be inaccessible because of transient network outages.
    \item 
    \textbf{Rule-resolvable errors}:
    As many multi-agent applications require information exchange between services or agents, it is essential to follow the protocols for those communications, e.g., in JSON format.
    However, as the responses of LLMs are not fully controllable yet, their return may not follow the format required in the prompts.
    For example, we may expect a response from an LLM in JSON, but a right brace is missed at the end of the return, leading to parsing failure.
    As the JSON format has clear specifications, it is reasonable to assume that a subset of these errors can be resolved by correcting the format according to the rules to meet the specifications.
    \item 
    \textbf{Model-resolvable errors}:
    When a multi-agent system handles some complicated tasks, the ability of the agent to understand the input, make decisions, and deliver outputs mostly depends on the capability of LLMs.
    In some cases, the responses of LLMs are in the expected format, but the content has problems, such as argument errors, semantic errors, or programming mistakes.
    It is hard to have pre-defined rules to regularize those responses for diverse tasks, but it has also been shown that such errors may be detected and recovered by further interaction with the LLMs.
    \item 
    \textbf{Unresolvable errors}:
    Eventually, there must be some errors that cannot be detected or solved.
    A typical example is that the API key of an LLM is expired or unauthorized.
    The agents relying on it or the system can do nothing to resolve such errors without human intervention.
\end{itemize}

\paragraph{Fault Tolerance mechanisms in \ours}
In \ours, we provide different mechanisms to encounter the errors summarized above.
\begin{itemize}
    \item 
    \textbf{Basic auto-retry mechanisms}:
    To combat accessibility errors, \ours's API services and model wrappers are fortified with retry logic that developers can customize, such as setting the maximum retry count. This ensures that agents can recover from sporadic disruptions and maintain their operational continuity.
    \item 
    \textbf{Rule-based correction tools}:
    The rule-based correction tools are introduced into \ours to efficiently and economically handle some easy-to-fix format errors in the responses of LLMs.
    For example, we establish a set of default rules in \ours that can complete unmatchable braces and extract JSON data from strings. 
    Such rule-based correction tools can correct some of the common rule-resolvable errors without calling LLM APIs again, which means shorter processing time and no LLM API call cost.
    \item 
    \textbf{Customizable fault handlers}:
    \ours also integrates flexible interfaces of fault handlers in model wrappers for developers to define how to parse the responses from LLMs and handle the unexpected outputs.
    Application developers can configure their fault handling mechanism by providing a parsing function, fault handling function, and the number of chances given to LLMs through configurable parameters (i.e., \texttt{parse\_func} and \texttt{fault\_handler} and \texttt{max\_retries}) when invoking LLMs.
    With such developer-friendly design, \ours can be configurably robust to rule-resolvable errors (when the build-in rules fail to handle) and some model-resolvable errors that can be detected and handled by a single agent (e.g., distilling a verbose summary to a more concise one).
    \item 
    \textbf{Agent-level fault handling}:
    There are model-resolvable errors that require more advanced LLM usages or agent-level interaction to recover.
    For example, detecting semantic errors, which usually include factual inaccuracy, logical inconsistency, contextual incoherence, unreasonable inference, and inappropriate vocabulary usage, is challenging since they may not necessarily trigger immediate red flags within the system's existing validation processes. 
    Developers can utilize the agent's ability in \ours (e.g., memory module and message hub) to critique for semantic error checking such as self-critique, pairwise critique, and human-augmented critique.
    \item 
    \textbf{Logging system}:
    Although the unsolvable errors are too tricky for the system to handle, \ours provides an improved logging system for developers to quickly monitor and identify the problems in multi-agent applications.
    The logging system in \ours has customized features for the multi-agent application scenarios, including adding a logging level called \texttt{CHAT} for logging conversations between agents, providing formatted logs with various execution information, and a WebUI user interface to facilitate monitoring.
\end{itemize}

\section{Multi-Modal Applications}
\label{sec:mutil-modal}

The integration of multi-modal data is indispensable for advancing the capabilities and applications of multi-agent with LLMs.
\ours is designed to seamlessly support various data modalities, leveraging the diverse inputs and outputs that contemporary LLMs can process and produce.

\begin{figure}[ht]
    \centering
    \includegraphics[width=0.7\linewidth]{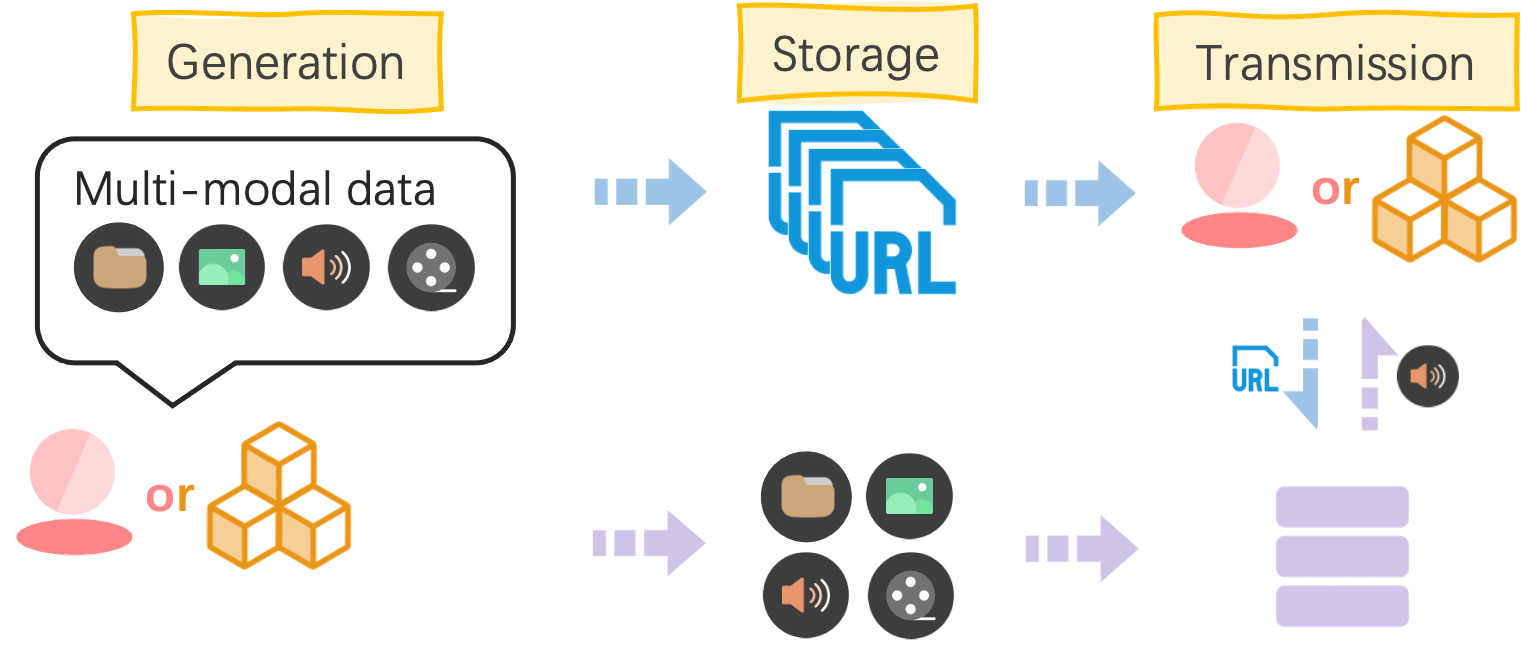}
    \caption{The generation, storage, and transmission of Multi-modal data in \ours.}
    \label{fig:modals}
\end{figure}

\paragraph{Management of Multi-Modal Data}
In a running \ours application, the lifecycle of multi-modal data is carefully managed. This management includes the generation, transmission, and storage of multi-modal data—all facilitated through a decoupled architecture using URLs and a local file manager system. 
\figref{fig:modals} exemplifies this process, including data originating from user inputs or model generations, data storage and retrieval, and data sharing.

\begin{itemize}
    \item 
    \textbf{Multi-modal data generation}:
    There are two primary sources of multi-modal data in \ours.
    One source is simply the locally stored multi-modal files, which can be used by either user proxy agents or general agents with access to the local file system.
    Another source is the model-modal content generation models.
    Our model APIs and the model wrappers integrate the most popular multi-modal models, such as the text-to-image content generation models like OpenAI's DALL-E, and conversely, the image-to-text image analysis models, e.g., GPT-4V.
    Besides the built-in APIs, developers can introduce their favorite multi-modal models and customize their own model wrappers, with our ready-to-use examples as the starting points.
    This customization process is streamlined in \ours and benefits from our modular design, allowing developers to connect their multi-modal services with minimal effort.
    \item 
    \textbf{Multi-modal data storage}:
    As mentioned above, multi-modal data in the multi-agent application can be either from ready-to-use local files or generated by multi-modal models.
    When a multi-modal model wrapper is invoked to generate multi-modal data, it first saves the data locally with the help of the file manager and returns a local URL when it receives multi-modal data from the model API service. 
    \item
    \textbf{Multi-modal data transmission}:
    \ours simplifies the process of multi-modal data sharing between agents by allowing agents to encapsulate local or remote URLs in multi-modal messages to indicate the actual storage locations of the data.
    The receiver agents can load the multi-modal data through the URLs when ready to process those.
\end{itemize}

The benefits of introducing URLs in the messages when agents share multi-modal data are three-fold.
Firstly, it can minimize the message size to avoid potential errors or delays because of the network bandwidth and enable the receiver agent to load the data on demand.
Secondly, if there is other text information in the message, the downstream agents can potentially prioritize or parallel the processing of the text information to/and the processing of multi-modal information.
Last but not least, such URL-attached messages can also facilitate the multi-modal data demonstration, which will be introduced in the following section.

\paragraph{Multi-Modal Interaction Modes}
With the implementation of URL-attached messages, \ours empowers users to interact with multi-modal systems via accessible interfaces such as terminal and web UI.  \figref{fig:modal_webui} showcases the user's ability to interact with multi-modal data within interaction modes. 
In the terminal, users can conveniently access locally stored data by activating the provided URLs. The web UI further enhances user experience by providing an intuitive platform to view and analyze multi-modal content, aligning with the expectations of modern web applications.

\paragraph{} Through \ours, developers are equipped to tailor model API services and wrappers to their individual needs, forge applications that handle diverse data modalities, and provide users with the necessary tools to engage with multi-modal agents effectively. This comprehensive support for multi-modal applications positions \ours as a versatile and powerful framework for harnessing the full potential of multi-agent LLMs, broadening the horizons for developers and researchers alike in creating sophisticated and interactive AI systems.

\section{Tool Usage}
\label{sec:tools}
Tool usage is an important feature for LLM-empowered agents, allowing agents to perceive, change their environment, and handle more complex tasks~\citep{auto-gen,art,talm}. 
For simplicity, we treat using tools as equivalent to calling service functions by LLMs. 
In \ours, the tool usage module is designed based on ReAct algorithm~\citep{react}, which allows for the generation of interleaved reasoning and task-specific actions, along with a core component---service toolkit. 
Such design features high compatibility, extensibility, robustness, and re-usability, spanning from function pre-processing, prompt engineering, reasoning, and response parsing to agent-level fault tolerance. 
Specifically, in \ours the tool usage involves four steps:

\begin{itemize}
	\item \textbf{Function Preparation}: 
	Parse the provided service functions, and pre-process the functions so that LLMs can utilize them directly. 
	
	\item \textbf{Instruction Preparation}: 
	Prepare instruction prompt for tool usage to elaborate the available tool functions to LLMs, including the purpose, arguments, constraints of the function, and its calling format. 
	
	\item \textbf{Iterative Reasoning}: 
	LLMs generate strategic reasoning, make decisions for tool usage, and respond in the required format. 
	
	\item \textbf{Iterative Acting}: 
	Parse and check the LLM response according to the calling format, invoke functions if the response adheres to the expected format, or generate a detailed error message to LLMs for correction. 

\end{itemize}

\begin{figure}[t]
\centering
	\includegraphics[width=0.65\linewidth]{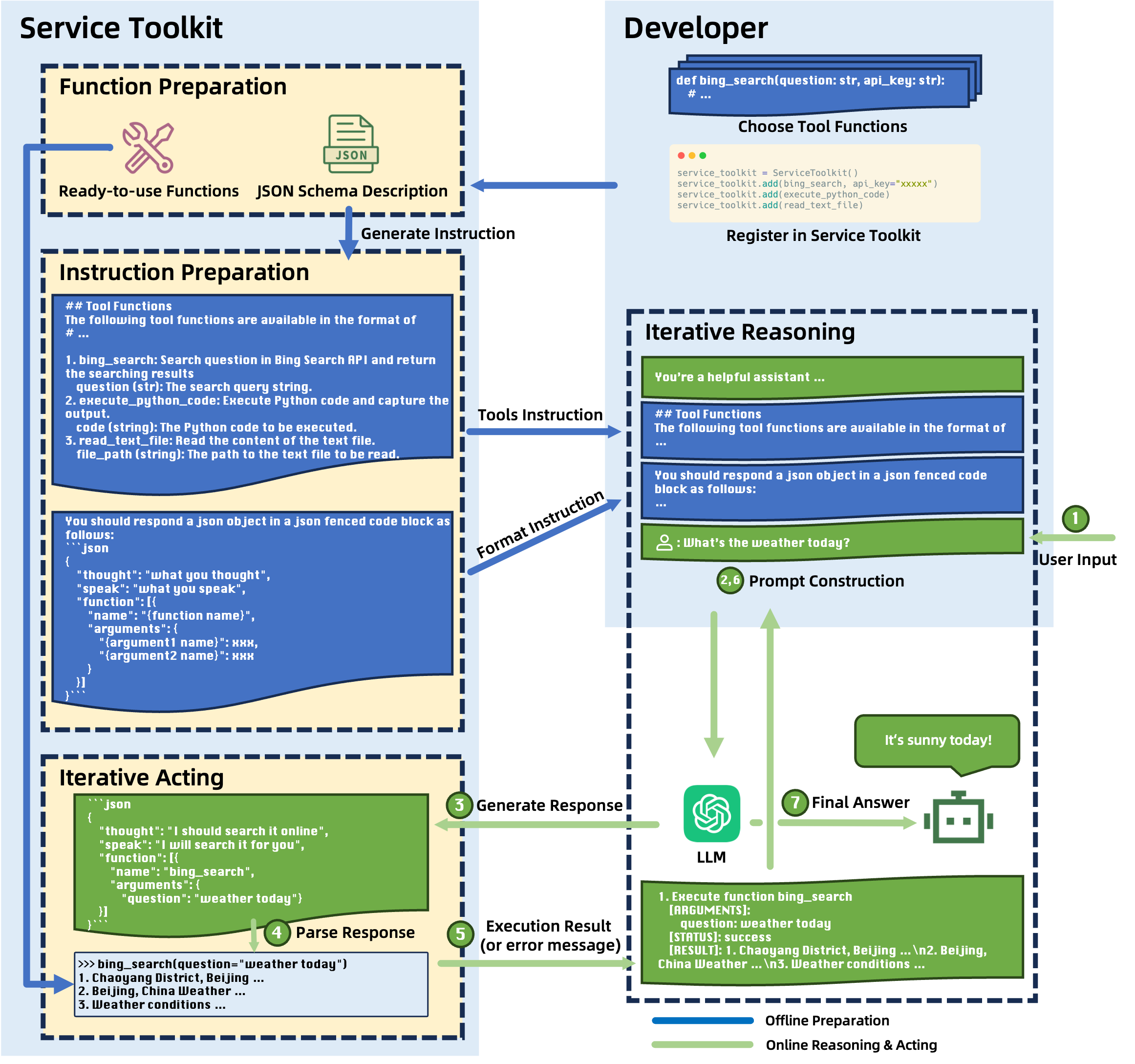}
	\caption{The ReAct-based tool usage module in \ours.}
	\label{fig:tools}
\end{figure}

In the above process, the service toolkit module is responsible for tool functions management, pre-processing, prompt engineering, response parsing, and function execution, and it is highly modular and extensible. 
\figref{fig:tools} demonstrates how the service toolkit works in \ours when users post a query.

\textbf{Function Preparation}.
In function preparation, the target is to preset the developer-specific arguments, and to generate ready-to-use functions and their corresponding formatted description for LLMs.
In \ours, developers only need to register their functions with preset arguments in the service toolkit. 
As shown in \figref{fig:tools}, developers choose the Bing search function and provide the API key during registration.
Then the service toolkit will automatically generate the processed ready-to-use function and its description in JSON schema format. 
The descriptions will be used to generate tool instructions in natural language. 
Optionally, some model APIs (\eg OpenAI and DashScope Chat API, \etc) can receive the JSON schema descriptions directly, which we will discuss in \secref{sec:customization}.

\paragraph{Instruction Preparation}
For novice developers, the service toolkit builds in templates for tool instruction and calling format for tool usage, as demonstrated in \figref{fig:tools}. 
The tools instruction template lists each function with a clear description and the parameters it requires, leading to an easy understanding of their functionalities. 
On the other hand, the calling format, as demonstrated in \figref{fig:tools}, requires a JSON dictionary in a Markdown fenced code block with \textit{thought}, \textit{speak}, and \textit{function} fields. 
During LLM generation, we expect the \textit{thought} field will provide a reasoning process for the next acting, including analyzing the current situation, selecting candidate functions, and correcting errors. 

\paragraph{Iterative Reasoning}
In \ours the reasoning and acting steps are iterative. 
As stated above, in the reasoning step LLMs should analyze the current situation and decide the next actions. 
Developers only need to construct prompts with the tool instructions and the calling format instructions and feed them into the LLMs. 
Such design provides high reusability and flexibility, that is, the service toolkit is task-independent and can be adapted to different tasks and scenarios very easily. 

\paragraph{Iterative Acting}
In the acting step, the service toolkit will parse the LLM response according to the calling format, extract the selected function, and execute it with the corresponding arguments. 
If the response conforms to the format requirements, and the function executes successfully, the service toolkit will return the execution results directly, which LLM can generate a response based on in the next reasoning step. 
Otherwise, we break down errors into response parsing errors, function execution errors, and other runtime errors. 
For response parsing and function execution errors, we expose them to LLM with detailed error information for correction in the next reasoning-acting iteration, leaving the other runtime errors to developers. 

\subsection{Customization for Experienced Developers}
\label{sec:customization}

\ours supports developers in highly customizing their tool instructions and function calling formats. 
To customize tool instruction, the service toolkit in \ours provides JSON schema descriptions automatically, which provides a structured way to elaborate how a function should be called, including its name, purpose, arguments, and other relevant details. 
These formatted descriptions can be directly fed into some advanced model APIs, e.g. OpenAI and DashScope Chat APIs. 
For users who want to deeply customize their tool instructions, they can construct instructions based on the JSON schema descriptions. 

\begin{figure}[th]
	\centering
	\includegraphics[width=0.8\linewidth]{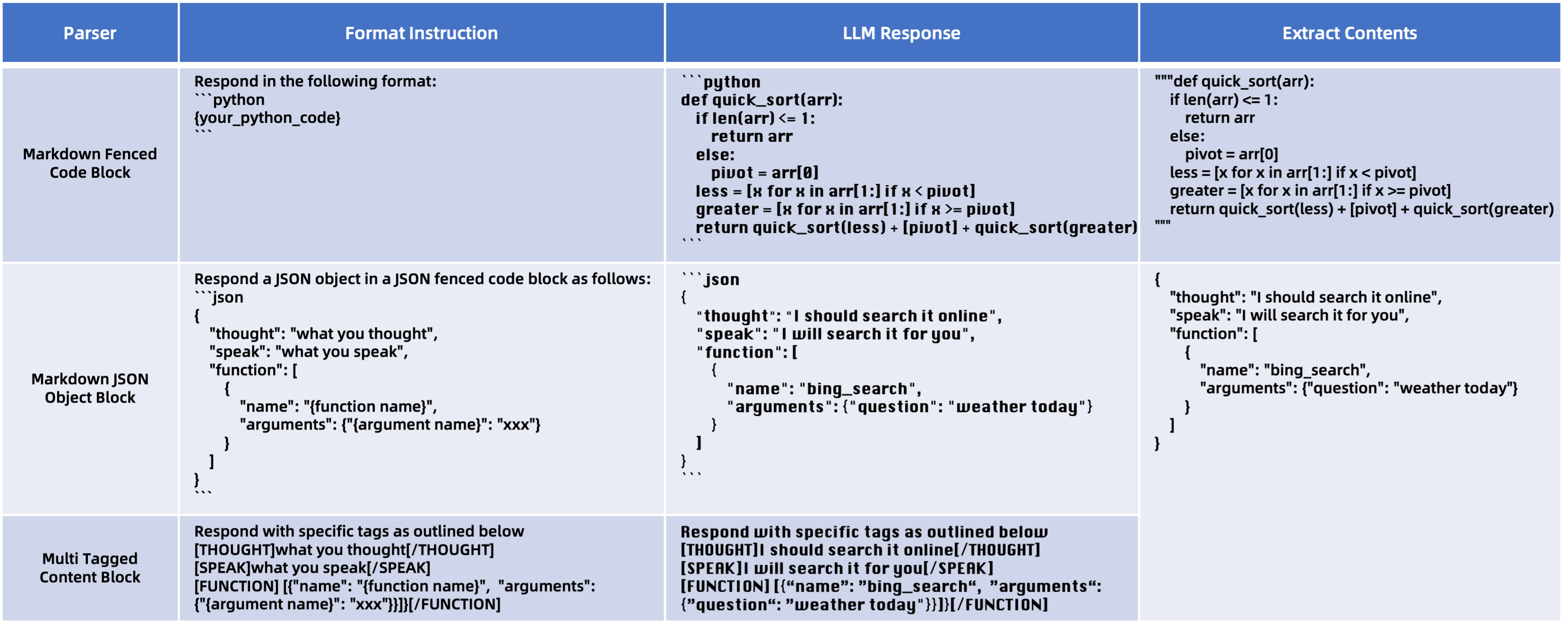}
	\caption{Parsers in \ours.}
	\label{fig:parsers}
\end{figure}

Besides the tools instruction, \ours also provides great flexibility, that is, \ours provides various model response parsers, including Markdown fenced code blocks, JSON object code blocks, and customizable tagged contents, as demonstrated in \figref{fig:parsers}. 
For the users who want to customize the function calling format, the Markdown fenced code blocks and JSON object code block allow them to quickly construct the format instruction and parse the LLM response according to the content types. 
For users who want to obtain multi-fields from LLMs, the multi-tagged contents allow the developers to combine different tagged contents at will and extract them easily from the response into a Python dictionary. 
With these parsers, developers are able to customize their own calling format easily. 

\section{Agents with Retrieval-Augmented Generation}
\label{sec:rag}

With the growing applications of LLMs, some circumstances require knowledge that is not contained in the training data set, for example, knowledge in highly professional domains or not publicly available. Even given the required datasets, the fine-tuning or re-training of the LLMs is still expensive. Accordingly, \textit{retrieval-augmented generation} (RAG), an innovative approach that aims to boost the power of LLMs in customized knowledge domain~\citep{rag_survey1,rag_survey2}, is gaining increasing attention in the literature.

The methodology of RAG can be considered as inserting a pre-processing step into the common utilization pipeline of LLMs. That is, given a collection of documents that contains needed knowledge, a similarity-based index is built, and the original user input is zipped with the most relevant pieces of information and converted into prompts, then fed to the LLMs. Therefore, the methodology of RAG involves multiple phases, that is, the collection of documents that contain the necessary information, the segmentation of the documents, the indexing of the segments (a.k.a. chunks or nodes), the similarity-based index retrieval, the fusion of the original query (i.e. user input) and retrieved results, the composing of prompts, and lastly, generation of reasonable responses from the LLM based on the informative prompts. 

In short, RAG embraces both the power of information retrieval and the generative capabilities of LLMs, and provides enhanced LLM service with customized domain knowledge at low cost. Meanwhile, assisted by RAG, the hallucination could be avoided and the factual accuracy could be significantly improved. 

As a developer-oriented multi-agent platform, \ours provides comprehensive RAG support for multi-agent applications. 
Given popular RAG frameworks such as \eg LlamaIndex~\citep{llamaindex}, LangChain~\citep{langchain}, \etc, \ours is designed with highly flexible abstracted processes to be compatible with those frameworks. In what follows, we introduce several key features of \ours RAG.

\paragraph{Configurations in One-Stop}
Due to the complexity of the working pipeline, the configuration of RAG services is highly convoluted and often headache-some for users. While the RAG service provided by \ours is comprehensive and also involves multi-agent workflow, \ours provides a simple one-stop configuration solution by using a single \texttt{.json} file to group all RAG-related configurations. 

With this highly systematized configuration interface, users only need to focus on constructing the workflow, without being distracted by the repetitious configurations. For example, the RAG-empowered agents may involve a wide collection of knowledge bases that need to be configured in detail. With this ``One-Stop'' feature, the corresponding adjustments of the modules (which may lead to different performances) are integrated as the editing over simply one single file. Moreover, this solution also naturally adapts to the \ours Workstation, in which the dialog-box-based configuration can be easily exported to executable files and later loaded in Python programs.


\paragraph{Knowledge-oriented Data Managements}
The application of RAG in multi-agent circumstances is more complicated compared with the application on a single agent. For example, for a single agent, one can directly encapsulate the needed knowledge to the agent. Therefore, the initialization of each RAG agent involves the whole pipeline of conversion from the original documents to vector-stored indexes with retrievers. However, in multi-agent applications, it is natural for agents to share knowledge, such that repeatedly executing index computation for each agent is needless. Therefore, \ours introduces the notion of \texttt{knowledge banks}.

Knowledge banks can be considered as a collection of knowledge containers, where the smallest manageable unit is a customized object (which will be referred to as a ``RAG object'' in the following context). The workflow starts with initializing the knowledge bank, which mainly relies on the information contained in the \texttt{.json} configuration file. The information includes the directory and extensions (such as \texttt{.py} or \texttt{.md}) of documents, the granularity and choice of segmentation tools (e.g. the splitters in \texttt{Llama-Index}) for the documents, and choice of model for indexing. After the initialization, the computed results are persisted to the designated directory for later use and we also obtain a knowledge bank consisting of RAG objects, each marked with a unique \texttt{knowledge\_id}, associated with the index of the corresponding documents, an information retriever, and other attributes. Note that \ours permits each RAG agent to load with more than one RAG object.


\paragraph{Agents with RAG}

The application of agents with RAG in \ours is very simple. For example, we first need to initialize a \texttt{KnowledgeBank} with some RAG framework, e.g. LlamaIndex, and all the documents. Then, we configure an RAG agent and load it with the knowledge bank. After that, the initialization is completed and we can use the RAG agent like any other agent in \ours. It is worth noting that if \texttt{KnowldgeBank} is obtained with LlamaIndex framework, then we need to use \texttt{LlamaIndexAgent} (inherited from \texttt{RAGAgentBase}). The readers may refer to Section~\ref{sec:copilot} for a concrete application sample, which implements a copilot for \ours using our RAG agents. Overall, the key features of RAG agents are summarized as follows:
\begin{itemize}
    \item The RAG agent is permitted to load several RAG objects (i.e. any subset of the knowledge bank). One can choose to load the original RAG objects from the knowledge bank (in such case, the modification to an object may affect all the agents who use it) or a copy of it.
    \item While agents are initialized with a \texttt{KnowledgeBank} object, it is permitted for the agents to update knowledge in time. The operations include inserting, deleting, or replacing knowledge pieces. Moreover, we provide a solution by monitoring certain directories and keeping the RAG object updated with the contents in the directories.
    \item The fusion mechanism of the retrieved results from multiple RAG objects is fully customizable. For example, since knowledge may be of different importance or trustworthiness, the agent can set weights for information retrieved from different RAG objects for subsequent processes. 
    \item RAGs agents are permitted to recompose the query in configurable repeats and conduct multiple queries for more comprehensive answers.
\end{itemize}
\section{Actor-based Distributed Framework}
\label{sec:distribute}

Efficiency and extensibility are essential when building industry-level applications on multi-agent systems.
The inference speed of the agents in multi-agent applications may vary dramatically. 
For example, suppose an agent in a multi-modal application employs a text-to-video model. In that case, its response time may be significantly longer than that of an agent designed to fill in details of stories.
\emph{Parallelization}, as a classic idea, should be introduced to boost efficiency.
Besides, multi-agent applications can comprise agents physically distributed on different machines.
A typical use case is that a company can wrap its patented techniques or private knowledge bases into an agent on their local machines connected to the internet and provide autonomous services to other entities via agent interactions.

However, when it comes to multi-agent systems, a challenge is that developers need to make decisions between the following two pairs of technology roadmaps.
As there is no free lunch, any combinations have their benefits and drawbacks. 

\begin{itemize}
    \item 
    \textit{Centralized v.s. decentralized coordination.}
    In the context of the distributed system, centralized coordination means multiple computation nodes being managed by a central node, such as the server-client model.
    A multi-agent mechanism with centralized coordination means the execution of the agents is scheduled by, and the messages between agents are forwarded by a central coordination component.
    On the contrary, decentralized coordination does not rely on any central component to schedule or forward messages, but the agents in such a system can be invoked automatically and send messages directly to the downstream agents for further processing.

    While centralized coordination is a straightforward style that can be understood and is easy to debug, its disadvantages include vulnerability to central node failures, imposing heavy traffic on the central node, and difficulty in scaling or extending to complicated applications.
    In contrast, decentralized coordination may require extra effort to develop and maintain but has a higher robustness against failure of any single node.

    \item 
    \textit{Static vs. dynamic workflow design.}
    A similar comparison can be found between the static computational graph employed in early versions of TensorFlow~\citep{abadi2016tensorflow} and the dynamic computation graph used in PyTorch~\cite{paszke2019pytorch}.
    In the context of multi-agent applications, the choice between a static and dynamic workflow is akin to choosing between pre-compiled and interpreted execution.
    The static workflow design can enable the optimization of the workflow graph level for running time and resource allocation.
    However, static workflow design requires the workflow graph to be known before execution, which limits the adaptation into applications, especially the ones with loop structures in the design.
    In contrast, dynamic workflows offer greater flexibility at the expense of optimization potential. This is particularly relevant when dealing with large language models where execution paths can change based on the input data or model inference results.
\end{itemize}

\begin{figure}[th]
    \centering
    \includegraphics[width=0.65\linewidth]{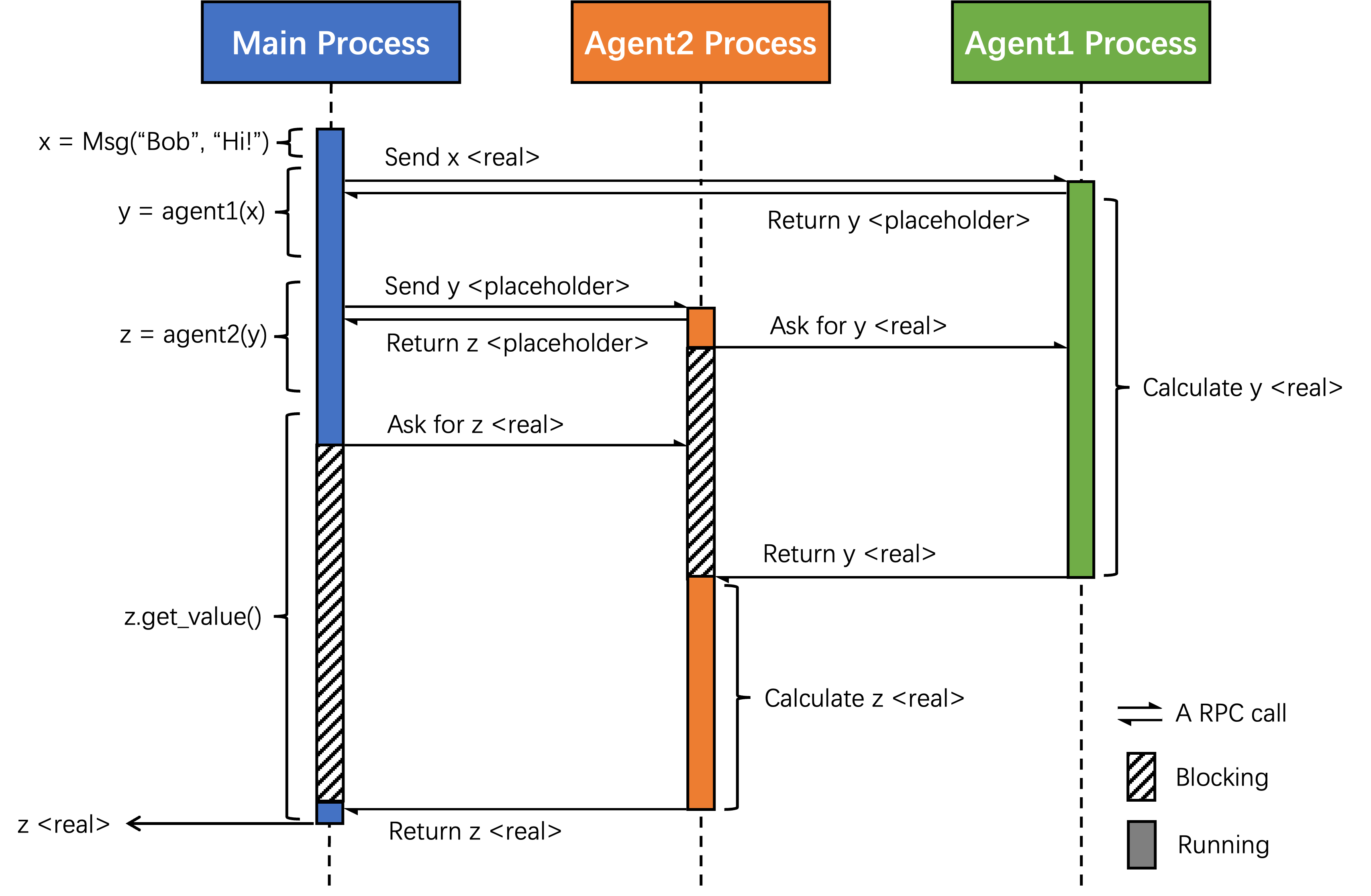}
    \caption{An example of a distributed application in \ours, illustrating various processes as denoted by different colors.}
    \label{fig:actor}
\end{figure}

\paragraph{Distributed mode in \ours.}
\ours balances these technology roadmaps by implementing an actor-based distributed mode that is mindful of the unique needs of multi-agent LLM systems, with the following important features:

\begin{itemize}
    \item \textit{Automatic parallel optimization without static graphs.}
\ours leverages the actor model to enable automatic parallel optimization, allowing developers to circumvent the intricacies of static graph programming. This approach seamlessly aligns with the dynamic and often unpredictable nature of LLMs, where the computational graph can alter based on evolving contexts and dialogue states.
    \item \textit{Programming workflows with minimal complexity.}
In contrast to traditional actor models and peer-to-peer (P2P) implementations that require intricate execution ordering for distributed agents, \ours simplifies workflow programming to a single procedural style within a Python function. This design significantly flattens the learning curve for developers, making the construction of sophisticated multi-agent LLMs more accessible.
    \item \textit{Hybrid local and distributed agent support.}
\ours's flexibility extends to supporting a hybrid mode where some agents operate locally while others are distributed. This feature is particularly beneficial when integrating LLMs with varying computational requirements, allowing for resource-intensive models to be distributed while less demanding agents remain local, all without the developer needing to differentiate between the two during implementation.
\end{itemize}

Specifically, we can concisely describe how \ours incorporates the actor model as the following. 
In this conceptual framework, an ``actor'' acts as a stand-alone entity that processes computation upon receipt of all necessary messages. This paradigm ensures that each agent, corresponding to an actor, only engages in computation once the required input messages are ready, thus achieving automatic parallel optimization.

However, the actor-model-based workflow presents a programming challenge: the variable (i.e., messages) passing between actors (i.e., agents) may be placeholders without any practical meaning at the beginning. 
To alleviate this, \ours introduces the ``placeholder'' message, a novel data structure that allows the main process to continue without blocking, while preserving the necessary information to retrieve real values later (\figref{fig:actor}). This mechanism is particularly advantageous for multi-agent LLM systems, where execution flow must adapt to the variable output of language models.

\begin{lstlisting}[language=Python, float=h, caption={ Demonstrating the use of placeholders in control flow within \ours.}, label={lst:control_flow}]
# set up distributed agent: agent1
...

input_msg = Msg("system", "Which agent should respond next, agent2 or agent3?")

# the variable choice is a placeholder
choice: placeholder = host_agent(input_msg)

if choice["content"] == "agent2":
    response = agent2()
elif choice["content"] == "agent3":
    response = agent3()
\end{lstlisting}

Another series of challenges arise when placeholders are used within control flow statements (e.g., if-else, loops) without their real values. 
An example is shown in \lstlistingname~\ref{lst:control_flow}, where a placeholder is required to make decisions.
In these circumstances, \ours temporarily blocks the process to retrieve its actual value, thus ensuring the continuity of the control flow.

The actor-based distributed mode in \ours not only provides automatic parallel optimization and simplifies the developer experience but also demonstrates high efficiency for distributed multi-agent LLM applications. It enables developers to focus on implementing agent logic, particularly the ``reply'' function, without concern for underlying distributed complexities. This streamlined approach to distributed multi-agent systems can advance the field of LLMs by making it easier to develop, run, and debug sophisticated and scalable multi-agent architectures.

\textbf{One-click deployment in \ours.}
To further ease the distributed deployment, \ours provides \textit{agent server} and a unified message center, named \textit{\ours Studio}. 

Specifically, the agent server is hold in remote machines, which receives requests from \ours applications, and initialize their required agents in the deployed machine automatically. 
That means, developers can set up agent instances remotely, without programming in different machines. 
Such feature provides high flexibility, especially for large-scale simulations, where a large number of agent instances will be set up in remote machines. 

For \ours Studio, it provides a unified display interface for distributed multi-agent applications, where messages from all distributed agents will be gathered and displayed in this studio, and allows developers to forward these messages to their own display interface. 
Besides, \ours studio supports agent servers management, that is, in this studio developers can check the deployment of distributed agents, open or close agent servers remotely. 
With this studio, developers can manage their applications much more easily.

\section{Signature Applications of \ours}
\label{sec:app}

As introduced in the previous sections, \ours is a multi-agent platform delicately designed for integrating and coordinating large-scale models in a user-friendly and fault-tolerant manner, and it is an ideal platform for a vast spectrum of applications. 
\ours can implement applications spanning from a simple single-agent vs. user dialog to complicated interactive multiplayer role-play games like \textit{werewolf}. 
Moreover, beyond centralized deployments, \ours can extend to distributed conversations that involve parallel operations across multiple machines. In this section, we look into several signature applications of \ours that persuasively illustrate the framework's outstanding and diverse capabilities. All examples referenced herein are accessible in our GitHub repository for community use and contribution.

\subsection{Dialog Agents: Basic Conversation}

The simplest yet most fundamental application of \ours is the \textit{basic conversation}, where the user directly interacts with the dialog agent. This application is an excellent starting point for fresh users of \ours to quickly capture the core message-passing mechanism in our framework.

The \textit{basic conversation} example demonstrates the usage of two fundamental built-in agents in \ours, the \texttt{UserAgent} and \texttt{DialogAgent}, which facilitate inputs from the user and the responses from LLMs, respectively. Normally, as illustrated in  \lstlistingname~\ref{lst:stand_conv_1}, the first step of all applications is the initialization, which is to load the model configurations (specified in the \texttt{model\_configs.json} file) through the \texttt{init} interface of \ours, which assigns the LLM-empowered agents with selected models. Currently, \ours is compatible with various platforms and APIs, including but not limited to standard OpenAI chat/embedding/DALL-E, HuggingFace, ModelScope, and a collection of locally hosted models with FastChat, vllm, and Flask. Moreover, the \texttt{init} interface also specifies detailed options such as file storage, logging, agent configures, \etc.
With all the configurations settled, it is ready to construct the conversation flow, i.e. the message-exchanging mechanism between the user/agents, which is an essential building block for all agent-based applications. In this workflow, the AI agent will always respond to the user's input, the conversation could form an endless loop until the user decides to opt-out.

To implement more sophisticated applications, \ours facilitates \texttt{pipelines}, which provide a well-structured and scalable framework for complex agent interactions (in terms of messages). As illustrated in \lstlistingname~\ref{lst:stand_conv_3}, we can implement the \textit{basic conversation} example with a sequential pipeline or loop pipeline. Readers may also refer to \appref{app:conv} for conversation history while running the demo codes.

\begin{lstlisting}[language=Python, caption={Code example of the basic conversation example.}, label={lst:stand_conv_1}, float=t]
import agentscope
from agentscope.agents import DialogAgent, UserAgent

# read model configs
agentscope.init(model_configs="./openai_model_configs.json")

# Create a dialog agent and a user agent
assistant_agent = DialogAgent(
    name="Assistant",
    sys_prompt="You are a helpful assistant",
    model="gpt-4"
)
user_agent = UserAgent()

# Basic version
x = None
while x is None or x.content != "exit":
    x = assistant_agent(x)
    x = user_agent(x)
\end{lstlisting}

\begin{lstlisting}[language=Python, caption={Pipeline-based implementation of the basic conversation example.}, label={lst:stand_conv_3}, float=h]
# Advanced version with sequential pipeline
from agentscope.pipelines.functional import sequentialpipeline
x = None
while x is None or x.content != "exit":
    x = sequentialpipeline([dialog_agent, user_agent], x)

# Advanced version with while loop pipeline
from agentscope.pipelines.functional import whilelooppipeline
x = whilelooppipeline(
    [assistant_agent, user_agent], 
    condition_func= lambda _, x: x is None or x.content!="exit", 
    x=None)
\end{lstlisting}

\subsection{Dialog Agents: Group Conversation with Mentions}

Beyond the basic conversation between a user and a single dialog agent, \ours supports group conversations. To improve the interactivity, we introduce the ``mentions'' feature, which allows the user or agent to call a specific agent by simply ``@\texttt{agent\_name}''. The ``mention'' feature is supported by applying the \texttt{filter\_agents} function, which screens the message and identifies if any agent is mentioned in the message content.

In this example, we first initialize the agents involved in the conversation as shown in \lstlistingname~\ref{lst:group_conv_1}. Here, the characteristics of the agents can be customized in the \texttt{agent\_config.json} file, e.g. using \texttt{sys\_prompt} to customize the reaction style or functionality of the agents. Also, we utilize the message hub (\texttt{msghub}) to facilitate message deliveries among a group of agents. The \texttt{msghub} allows the sharing of public information (e.g. an announcement) and permits agents to broadcast messages to all agents. The conversation would end if a timeout limit is reached, or the user types in ``exit''. 

\begin{lstlisting}[language=Python, caption={Code example of the group conversations.}, label={lst:group_conv_1}, float=h]
import agentscope

# Read model and agent configs, and initialize agents automatically
npc_agents = agentscope.init(
        model_configs="./configs/model_configs.json",
        agent_configs="./configs/agent_configs.json",
    )
user = UserAgent()
agent = list(npc_agents)+[user]
...
# We use msghub to coordinate the conversations, ``hint'' is a message notified to all agents
with msghub(agents, announcement=hint):
while True:
    try:
        x = user(timeout=USER_TIME_TO_SPEAK)
        if x.content == "exit":
            break
    except TimeoutError:
        x = {"content": ""}
        logger.info(
            f"User has not typed text for "
            f"{USER_TIME_TO_SPEAK} seconds, skip.",
        )
    # if user mentions any npc_agent in the message, it will be added to the speak_list
    speak_list += filter_agents(x.get("content", ""), npc_agents)
    
    # if the speak_list is non-empty, the mentioned agents will respond in a sequential manner
    if len(speak_list) > 0:
        next_agent = speak_list.pop(0)
        x = next_agent()
    # otherwise, all agents will respond one by one.
    else:
        next_agent = select_next_one(npc_agents, rnd)
        x = next_agent()
    # if the response mentions any agent, it will be added to the speak_list
    speak_list += filter_agents(x.content, npc_agents)
\end{lstlisting}

\subsection{Dialog Agents: The Werewolf Game}

Group conversation and the mentioning feature are fundamental building blocks for multi-agent applications. Here we present a more sophisticated application, the \textit{werewolf game}, which is a popular multiplayer interactive role-play game. We aim to implement the game with \ours in only one hundred lines of code. This example involves six players divided into two opposing teams, the werewolves, and the villages. After rounds of conversations and discussions, the game ends when all werewolves are eliminated (i.e. villager victory) or the number of werewolves equals or outnumbers the villagers (i.e. werewolf victory).

As an LLM-empowered role-play game, we start the game settings with allocation for the roles and initialization for the agents. As shown in \lstlistingname~\ref{lst:wolf-init}, \ours supports a quick setup, which consists of default agent configurations for a user to instantiate the agent objects with corresponding roles in one click, the detailed settings are included in the \texttt{agent\_configs.json} file.

It is worth noting that the \textit{werewolf game} is based on the group conversation capability of \ours, such that the werewolves could chat in the ``night phase'' and all participants could discuss during the ``day phase''. Similar to the \textit{group conversation} example, the \textit{message hub} (\texttt{msghub}) of \ours is used to facilitate the conversations.  As shown in \lstlistingname~\ref{lst:wolf-init}, after the host (moderator) makes an announcement, the werewolves discuss for at most \texttt{MAX\_WEREWOLF\_DISCUSSION\_ROUND} rounds and conclude once an agreement is reached. Here, the agents are required to use an ``agreement'' attribute in the response message, which is enforced in the role-defining prompt. For complete workflow, an example of dialogue history, and more related information, please refer to \appref{app:wolf}.

\begin{lstlisting}[language=Python, caption={Code example of the werewolf game.}, label={lst:wolf-init}, float=t]
import agentscope
# Read model and agent configs, and initialize agents automatically
survivors = agentscope.init(
    model_configs="./configs/model_configs.json",
    agent_configs="./configs/agent_configs.json",
)

# Define the roles within the game.
roles = ["werewolf", "werewolf", "villager", "villager", "seer", "witch"]

# Based on their roles, assign the initialized agents to variables.
wolves, villagers, witch, seer = survivors[:2], survivors[2:-2], survivors[-1], survivors[-2]
...
# Night phase: werewolves discuss
hint = HostMsg(content=Prompts.to_wolves.format(n2s(wolves)))
with msghub(wolves, announcement=hint) as hub:
    ...
    for _ in range(MAX_WEREWOLF_DISCUSSION_ROUND):
        x = sequentialpipeline(wolves)
        if x.agreement:
            break
    ...
\end{lstlisting}

\subsection{Distributed Deployed Agents}

We have seen applications regarding conversations involving dialog agents, but those examples are fundamental in the sense that the agents are deployed in a centralized manner, that is, the agents are hosted on a single machine and in a single process. To allow agents to be hosted by separate machines or processes, \ours allows agents to be distributedly deployed in two modes, the single-machine multi-process mode, and the multi-machine multi-process mode. In what follows, we present examples to demonstrate this feature.

\noindent\textbf{Single-Machine Multi-Process Mode}: For this mode, all agents are deployed on a single machine, but running in separate processes. For better comparison, we implement the \textit{basic conversation} example in this mode (see \lstlistingname~\ref{lst:one_machine} for the complete code). Compared with \lstlistingname~\ref{lst:stand_conv_1} and \ref{lst:stand_conv_3}, we use the \texttt{to\_dist} function to convert the current agent instance into a distributed version. Then, the \texttt{assistant\_agent} would be deployed on a local host with an automatically allocated port. Besides the aforementioned differences, the single-machine multi-process mode is identical to local deployment, yet it has been optimized for parallel execution.

\noindent\textbf{Multi-Machine Multi-Process Mode}: To demonstrate this mode, we initiate the agent service (a \texttt{DialogAgent}) on a remote machine (as shown in \lstlistingname~\ref{lst:remote_launch}), and constructs a workflow (as shown \lstlistingname~\ref{lst:multi_machine}). One may note that the only difference comparing to the local deployed mode is that the agent server needs to be connected using specified URLs and ports before establishing the workflow.

Overall, for \ours, we can smoothly convert from the local deployment mode to the distributed mode and vice versa, with only minimal changes to the agent configuration and no modification to the workflow.

\begin{lstlisting}[language=Python, caption={Example that deploys agents in single-machine multi-process mode.}, label={lst:one_machine}, float=t]
from agentscope.agents import UserAgent, DialogAgent
import agentscope
# we use .to_dist() to convert the agent to distributed mode.
assistant_agent = DialogAgent(
    name="Assistant",
    sys_prompt="You are a helpful assistant",
    model="gpt-4"
).to_dist()
user_agent = UserAgent()

x = None
while x is None or not x.content != "exit:
	x = sequentialpipeline([assistant_agent, user_agent], x)
\end{lstlisting}

\begin{lstlisting}[language=Python, float=t, caption={Deploying a remote agent in multi-machine multi-process mode..}, label={lst:remote_launch}]
from agentscope.agents.rpc_agent import RpcAgentServerLauncher
from agentscope.agents import DialogAgent

# load model configurations
agentscope.init(model_configs="configs/model_configs.json")
# set server for the remote agent
server_launcher = RpcAgentServerLauncher(
    agent_class=DialogAgent,
    agent_kwargs={
        "name": "Assitant",
        "sys_prompt": "You are a helpful assistant.",
        "model": "gpt-4"
    },
    host="xxx.xxx.xxx.xxx",
    port=12010,
)
# start the server
server_launcher.launch()
server_launcher.wait_until_terminate()
\end{lstlisting}

\begin{lstlisting}[language=Python, caption={Example of setting sub-processes for agents in multi-machine multi-process mode.}, label={lst:multi_machine}, float=t]
agentscope.init(model_configs="configs/model_configs.json")
    
assistant_agent = DialogAgent(
    name="Assistant",
    model="gpt-4"
).to_dist(
    host="xxx.xxx.xxx.xxx",     # The target URL of agent server
    port=12010,                 # The target port of agent server
    launch_server=False,        # Use the remote agent server
)
user_agent = UserAgent()

x = None
while x is None or not x.content != "exit":
	x = sequentialpipeline([assistant_agent, user_agent], x)
\end{lstlisting}

\subsection{RAG Agents: \ours Copilot}\label{sec:copilot}

As previously introduced in Section~\ref{sec:rag}, Retrieval-Augmented Generation (RAG) allows developers to fully utilize the language generation capability of LLMs accompanied by a customized knowledge pool. Accordingly, \ours introduces RAG agents to facilitate such a functionality. In the following example (as shown \lstlistingname~\ref{lst:rag_agents_1}), we show how to use a collection of Llama-index-based RAG agents (i.e., the \texttt{LlamaIndexAgent} inherited from the \texttt{RAGAgentBase}) to build a multi-agent copilot for \ours.

We first initialize the agents. Note that the most important feature of RAG agents is that beyond customized personalities and behavioral styles configured by the system prompts, each agent is loaded with external knowledge, which is specified in the \texttt{agent\_configs} that contains configuration information such as data storage directory, targeted file types, document chunking settings, the indexing and embedding settings, \etc.

The workflow of \textit{copilot} is designed as follows, the user first inputs a message, and if the user mentions some specific RAG agents as we defined, then the corresponding agents would respond, otherwise, the \texttt{guide\_agent} would decide the most suitable agent to respond to the query. Due to space limit, we only represent simplified codes here and please refer to the repository and documentation for more details.

\begin{lstlisting}[language=Python, caption={Example of using RAG agents to build a copilot for \ours.}, label={lst:rag_agents_1}, float=t]
import agentscope
from agentscope.agents import UserAgent, DialogAgent, LlamaIndexAgent
...
# initialize agentscope with model configurations
agentscope.init(model_configs="configs/model_configs.json")

# initialize the RAG agents based on different configurations
tutorial_agent = LlamaIndexAgent(**agent_configs[0]["args"])
code_agent = LlamaIndexAgent(**agent_configs[1]["args"])
api_agent = LlamaIndexAgent(**agent_configs[2]["args"])
search_agent = LlamaIndexAgent(**agent_configs[3]["args"])
...
# initialize a basic dialog agent as the ``frontdesk assistant'' and a user agent
guide_agent = DialogAgent(**agent_configs[4]["args"])
user_agent = UserAgent()
...
while True:
    x = user_agent()
    # the workflow terminates when user inputs nothing or ``exit''
    if len(x["content"]) == 0 or str(x["content"]).startswith("exit"):
        break
    # find out the agents mentioned in user's input
    speak_list = filter_agents(x.get("content", ""), rag_agent_list)
    if len(speak_list) == 0:
        # if no agent is mentioned, the guide agent will decide which one to call
        guide_response = guide_agent(x)
        speak_list = filter_agents(
            guide_response.get("content", ""),
            rag_agent_list,
        )
    # agents called by the guide agent will be recorded
    agent_name_list = [agent.name for agent in speak_list]
    # the listed agents respond to the query in turn
    for agent_name, agent in zip(agent_name_list, speak_list):
        if agent_name in rag_agent_names:
            agent(x)
\end{lstlisting}

\subsection{Web Search and Retrieve Agents}

We have shown examples of agents generating responses by the capability of LLM (\texttt{DialogAgent}) and information retrieved from external knowledge libraries (\texttt{LlamaIndexAgent}). Nevertheless, we can also utilize internet resources to build agents, as introduced in the following example.

As presented in \lstlistingname~\ref{lst:search_01}, the initialization involves three types of agents - the \texttt{UserAgent} that takes user inputs, the \texttt{SearcherAgent} that converts the user's questions into keywords and calls the search engine to retrieve webpages from the internet, and the \texttt{AnswererAgent} that retrieves information from web pages to compose answers.

It is worth noting that, since a large number of web pages may be returned by the search agents. In the standard single-process mode, multiple \texttt{AnswererAgent} instances can only perform web searching and answer questions in a sequential manner on a single machine. For better efficiency, it is beneficial to allow multiple instances of \texttt{AnswererAgent} running in parallel, that is, the multi-machine multi-process mode of \ours agents. 

\begin{lstlisting}[language=Python, caption={Example of utilizing web search and retrieve agents.}, label={lst:search_01}, float=t]
import agentscope
from searcher_agent import SearcherAgent
from answerer_agent import AnswererAgent
from agentscope.agents.user_agent import UserAgent

agentscope.init(model_configs="configs/model_configs.json")

# we can perform multiple searches at one time
WORKER_NUM = 3
searcher = SearcherAgent(
    name="Searcher",
    model_config_name="my_model",
    result_num=args.num_workers,
    search_engine_type=args.search_engine,
    api_key=args.api_key,
    cse_id=args.cse_id,
)
# instantiate the answerer agents
answerers = []
for i in range(args.num_workers):
    answerer = AnswererAgent(
        name=f"Answerer-{i}",
        model_config_name="my_model",
    )
    # if we want to put agents in distributed (parallel) mode
    if args.use_dist:
        answerer = answerer.to_dist(lazy_launch=False)
    answerers.append(answerer)
user_agent = UserAgent()

msg = user_agent()
while not msg.content == "exit":
    msg = searcher(msg)
    results = []
    for page, worker in zip(msg.content, answerers):
        results.append(worker(Msg(**page)))
    for result in results:
        logger.chat(result)
    msg = user_agent()
\end{lstlisting}

\subsection{ReAct Agents: Convert Natural Language to SQL Query}

Natural Language to SQL query (NL2SQl) is a classical yet challenging task in both database and natural language processing communities, which aims to convert human input questions in natural language into SQL queries. In the research community, there is a collection of works exploring the potential of LLMs in NL2SQL, and it would be very interesting to explore this task with LLM-empowered agents.

In \ours, we provide a special class of agents, the ReAct (reasoning and acting) agents. More specifically, we could create new service functions, by using the \texttt{ServiceToolkit} module, for the ReAct agents and corresponding LLMs. In this example, we try to equip the ReAct agent with a state-of-the-art NL2SQL algorithm, DAIL-SQL. 

As the first step (as shown in \lstlistingname~\ref{lst:react_01}), we need to initialize the model config and the SQL database, then initiate and provide the corresponding database path in \texttt{sqlite} file format. Here we generate the SQLite file using the provided SQL commands. You can also use the \texttt{.sqlite} format file directly.

Then, as shown in \lstlistingname~\ref{lst:react_02}, we define the tools for ReAct Agent to execute the SQL query. Namely, our agent should be able to generate the SQL query given the natural language input and execute the SQL query to get the result. We referenced a third-party Text-to-SQL tool  \texttt{DAIL-SQL} to generate a Text-to-SQL prompt. We use the \texttt{query\_sqlite} service function in the \texttt{agentscope.service} module. Now, we can initiate the ReAct Agent using the defined tools and interact with the agent, as shown in \lstlistingname~\ref{lst:react_03}.

\begin{lstlisting}[language=Python, caption={Example of utilizing web search and retrieve agents.}, label={lst:react_01}, float=t]
import agentscope
from sql_utils import create_sqlite_db_from_schema
...
agentscope.init(model_configs="configs/model_configs.json")
create_sqlite_db_from_schema(db_schema_path, db_sqlite_path)
...
\end{lstlisting}

\begin{lstlisting}[language=Python, caption={Example of utilizing web search and retrieve agents.}, label={lst:react_02}, float=t]
from agentscope.service import (
    ServiceResponse, 
    ServiceExecStatus, 
    ServiceToolkit, 
    query_sqlite,
)
from sql_utils import DailSQLPromptGenerator

def generate_sql_query(question: str, db_path: str, model: Callable) -> ServiceResponse:
    prompt_helper = DailSQLPromptGenerator(db_path)
    prepared_prompt = prompt_helper.generate_prompt({"content": question})
    
    def get_response_from_prompt(prompt: dict, model: Callable) -> str:
        ...
    sql_response = get_response_from_prompt(
        prepared_prompt["prompt"], model=model
    )
    return ServiceResponse(
        ServiceExecStatus.SUCCESS,
        sql_response
    )

# Use Service Toolkit to set up tool functions for LLMs
service_toolkit = ServiceToolkit()
service_toolkit.add(generate_sql_query, db_path=db_sqlite_path, model=loaded_model)
service_toolkit.add(query_sqlite, database=db_sqlite_path)
\end{lstlisting}

\begin{lstlisting}[language=Python, caption={Example of utilizing web search and retrieve agents.}, label={lst:react_03}, float=t]
from agentscope.agents import ReActAgent
agent = ReActAgent(
    name="assistant",
    model_config_name='gpt-4',
    service_toolkit=service_toolkit,
    sys_prompt="You are a helpful agent that preform SQL queries base on natual language instructions.",
    verbose=True, # set verbose to True to show the reasoning process
)
...
mss = Msg(
    name="user", 
    content="How many singers do we have?", 
    role="user"
)
logger.chat(mss)

sql_query_mss1 = agent(mss)
...
\end{lstlisting}

\subsection{\ours Workstation}

\begin{figure}[ht]
    \centering
    \includegraphics[width=0.86\linewidth]{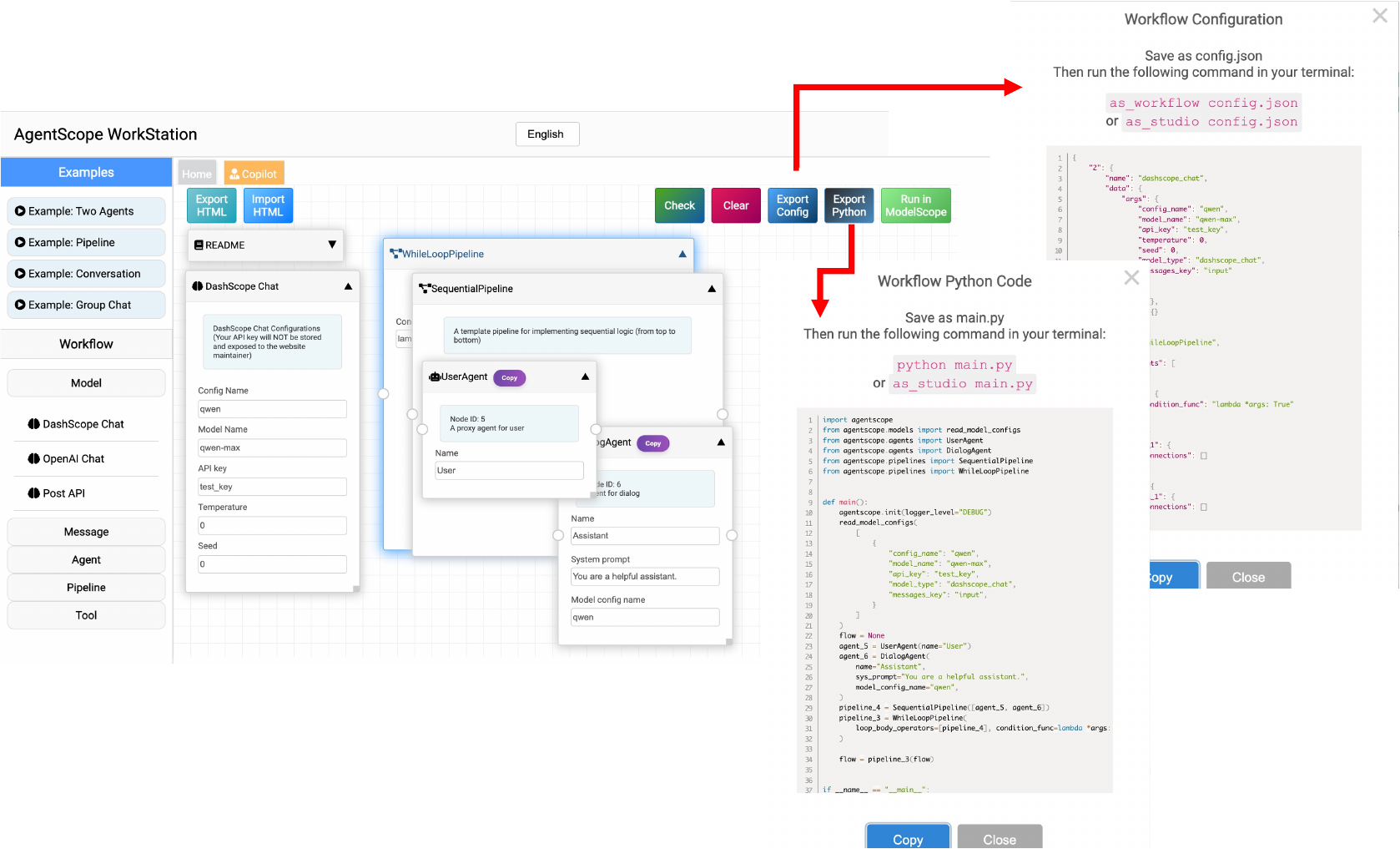}
    \caption{Workstation generates workflow configuration and Python code.}
    \label{fig:workstation_code}
\end{figure}

\ours provides a very convenient and user-friendly development kit in the form of ``dragging windows'', the \textit{Workstation}. Here, implementing applications of \ours using this development kit is of low cost in the sense that, entry-level developers or those without any programming experience could easily develop their own application at ease by simply dragging those agent-related modules and connecting them in a very straightforward way. For example, as shown in \figref{fig:workstation_code}, we implement the basic conversation example in \textit{Workstation}. As we can see, we do not need to write any code, just simply type in the configurations such as detailed settings and APIs into the corresponding windows, link the windows to build the dependency and connection, and then with one-click, \textit{Workstation} would get ready for launch automatically. Meanwhile, the \textit{Workstation} also introduced static checking rules to ensure the correctness of the configurations.

\ours Workstation also provides comprehensive support for advanced developers. The developers could export the configurations on the modules as \texttt{.json} files and execute by the \ours Workstation engine. Alternatively, one can also use the \ours Workstation Compiler to convert all configurations into Python codes for further editing or development to implement more customized adjustments.

\section{Related Works}
\label{sec:related_work}

The development of \ours aligns with the rapidly evolving landscape of frameworks that leverage large language models (LLMs) for the creation of language agents and multi-agent systems. Here, we briefly introduce works closely related to \ours from two sub-domains pertinent : Language Agent Frameworks, focusing on individual agent capabilities, and Multi-Agent Frameworks, emphasizing collaboration among multiple agents. For broader related works, readers can refer to \citep{agent-survey-1,agent-survey-2}.

\paragraph{Language Agent Frameworks}
Language agent frameworks are pivotal for developing applications that can interpret and interact using human language. 

The Transformers library \citep{transformers-agent} has introduced a natural language API to interface with transformer models in its recent updates (\textit{Transformers-Agents}). This API utilizes a set of customizable tools, allowing the model to interpret instructions and generate code snippets accordingly. It offers support for various open-source and proprietary model endpoints, catering to diverse developer needs.
\textit{LangChain} \citep{langchain} provides a framework for building applications that are context-aware and capable of reasoning. It includes libraries and templates that facilitate the integration of multiple components into a unified cognitive architecture. LangServe and LangSmith extend the framework's capabilities by enabling deployment as a REST API and offering developer tools for debugging and monitoring chains built on any LLM framework.
\textit{AutoGPT} \citep{autogpt} illustrates a different approach, allowing an LLM to iteratively execute actions and make decisions. As a generalist agent, AutoGPT is not task-specific; it is designed to perform a variety of computer-based tasks, reflecting the adaptive nature of LLMs.
\textit{ModelScope-Agent} \citep{modelscope-agent} is a customizable agent framework that harnesses open-source LLMs to perform tasks and connect with external APIs. It facilitates seamless integration with model APIs and common APIs while providing a comprehensive infrastructure for data collection, tool retrieval, and customized model training, all aiming to realize practical real-world applications.

\paragraph{Multi-Agent Frameworks}
Building on the capabilities of individual agents, multi-agent frameworks explore collaboration and interaction among multiple agents to address complex tasks.

\textit{AutoGen} \citep{auto-gen} provides a generic infrastructure that allows developers to program interaction patterns using both natural language and code. This framework enables the development of diverse applications by facilitating conversation among agents that are customizable and can utilize various combinations of LLMs, human inputs, and tools.
\textit{MetaGPT} \citep{meta-gpt} incorporates meta-programming to enhance multi-agent collaborations. By encoding Standardized Operating Procedures (SOP) into prompts, this framework ensures streamlined workflows and reduced errors, exemplifying effective task decomposition among agents.
\textit{AGENTS} \citep{Agents} is an open-source library that supports autonomous language agents with features like planning, memory, and multi-agent communication. It is designed to be user-friendly, helping non-specialists to deploy state-of-the-art language agents, and research-friendly, with a modularized design for extensibility.
\textit{OpenAgents} \citep{OpenAgents} provides an open platform for using language agents with practical functionalities accessible through a web interface. This framework emphasizes facilitating real-world agent interactions and includes specialized agents for different tasks, such as data analysis and web browsing.
\textit{ChatDev} \citep{ChatDev} exploits LLMs for software development, creating a virtual chat-powered company that follows a waterfall model. It engages ``software agents'' at different stages of the development process, facilitating collaboration and context-aware communication.
\textit{CAMEL} \citep{camel} proposes a novel framework for autonomous cooperation among communicative agents using role-playing techniques, which allows for the generation of conversational data for studying agent behaviors and capabilities.
Lastly, \textit{AgentSims} \citep{AgentSims} introduces a sandbox environment to evaluate LLMs in task-based scenarios, offering an infrastructure for researchers to test specific LLM capacities within a simulated environment.

These frameworks represent significant strides in the use of LLMs for both individual and collaborative agent tasks. \ours is situated within this context, contributing by addressing the need for a user-friendly, fault-tolerant and versatile framework designed to manage complex interactions and processes inherent in multi-agent LLM systems. By focusing on ease of use and reliability, \ours aims to facilitate the creation of robust and versatile applications across diverse domains.

\section{Conclusion}
\label{sec:conclusion}

In this work, we propose \ours, a platform that stands at the forefront of multi-agent system development, synergizing user-centric design with the advanced capabilities of LLMs. Through its innovative communication and distributed mechanisms, \ours demonstrates its potential to boost collaboration amongst agents, enabling efficient, fault-tolerant operations and multi-modal interactions. By abstracting complexities and offering an array of development utilities, \ours substantially lowers the barriers to entry, fostering a more inclusive and creative community of developers.

Looking forward, \ours opens numerous avenues for further research and development. Future work could delve into deeper integration of retrieval augmented generation, and explore adaptive communication protocols and interactive modals that evolve alongside task requirements. The platform's impact on accelerating the deployment of multi-agent systems across industries, from healthcare to customer service, promises to be profound, potentially leading to smarter and more responsive technologies that enhance human-machine collaboration.
With \ours, we invite the broader research and development community to build upon our foundation, driving innovations that will shape the next generation of intelligent multi-agent applications.

\clearpage
\bibliographystyle{plainnat}
\bibliography{citation}
\clearpage

\appendix

\section{Running Example of Standalone Conversation}
\label{app:conv}

\lstref{lst:conv_history} shows the dialog history when running the standalone conversation. 

\begin{lstlisting}[language=Dialog, float={th}, caption={A running example of standalone conversation.}, label={lst:conv_history}]
Assistant: Thank you! I'm here to help. How can I assist you today?
User: I wanna know about large language models. 
Assistant: Large language models are AI models trained on vast amounts of text data to understand and generate human-like text.
User: What's the most famous and powerful LLM in the world? 
Assistant: The most famous and powerful large language model currently is OpenAI's GPT-4.
User: What about the other LLMs?
Assistant: There are several other notable large language models including:

1. BERT (Bidirectional Encoder Representations from Transformers) by Google: This model is designed to understand the context of words in a sentence by looking at what's before and after them.

2. T5 (Text-to-Text Transfer Transformer) also by Google: This model treats every NLP problem as a text generation task.

3. RoBERTa by Facebook: This is a variant of BERT and is trained on more data and for a longer amount of time.

4. XLNet by Google Brain: This model overcomes some of the limitations of BERT by taking into account all possible word sequences in a sentence.

5. Turing-NLG by Microsoft: This is a 17-billion parameter language model by Microsoft, designed for various NLP tasks.

6. Megatron by NVIDIA: This is designed to train Transformer models with billions of parameters. 

These models are used for a variety of NLP tasks including translation, question answering, summarization, and more.
User: Okay, what should I know about LLM? 
Assistant: Large Language Models (LLMs) are powerful AI tools that can generate human-like text, but they require careful handling due to potential risks like generating harmful or biased content.
User: Okay, thanks.
Assistant: You're welcome! If you have any other questions, feel free to ask.
User: exit
\end{lstlisting}

\section{Running Example of Werewolf}
\label{app:wolf}

Please refer to https://github.com/modelscope/agentscope for the complete code of the werewolf game. We provid an example dialogue history in \lstref{lst:wolf_history}. 

\begin{lstlisting}[language=Werewolf, caption={Running example of werewolf.}, label={lst:wolf_history}]
Moderator: Player1 and Player2, you are werewolves. If you are alone, eliminate a player, else discuss with your teammates and reach an agreement. Response in the following format which can be loaded by python json.loads(){
    "thought": "thought",
    "speak": "thoughts summary to say to others",
    "agreement": "whether the discussion reached an agreement or not(true/false)"
}
Player1: I think we should consider Player3. They have a knack for figuring things out.
Player2: I agree with your point about Player3, they are indeed a strong player. But we should also consider Player4, who is also a strong player and could be the seer or witch.
Player1: I see your point about Player4. They could be a threat if they are the seer or witch. But I still think Player3 is a strong player and could figure us out.
Player2: Alright, let's go with Player3. We can consider Player4 next time.
Moderator: Which player do you vote to kill? Response in the following format which can be loaded by python json.loads()
{{
   "thought": "thought" ,
   "speak": "player_name"
}}
Player1: Player3
Player2: Player3
Moderator: The player with the most votes is Player3.
Moderator: Player6, you're witch. Tonight Player3 is eliminated. Would you like to resurrect Player3? Response in the following format which can be loaded by python json.loads()
{
    "thought": "thought",
    "speak": "thoughts summary to say",
    "resurrect": "true/false"
}
Player6: I'm deeply saddened by the loss of Player3. It's a tough decision to make.
Moderator: Player5, you're seer. Which player in Player1, Player2, Player3, Player4, Player5 and Player6 would you like to check tonight? Response in the following json format which can be loaded by python json.loads()
{
    "thought": "thought" ,
    "speak": "player_name"
}
Player5: Player1
Moderator: Okay, the role of Player1 is werewolf.
Moderator: The day is coming, all the players open your eyes. Last night is peaceful, no player is eliminated.
Moderator: Now the alive players are Player1, Player2, Player3, Player4, Player5 and Player6. Given the game rules and your role, based on the situation and the information you gain, to vote a player eliminated among alive players and to win the game, what do you want to say to others? You can decide whether to reveal your role. Response in the following JSON format which can be loaded by python json.loads()
{
    "thought": "thought" ,
    "speak": "thought summary to say to others"
}
Player1: I'm glad that we all made it through the night. I think we should keep an eye on Player4. They've been pretty quiet, which could mean they're trying to avoid suspicion.
Player2: I agree with Player1. Player4 has been unusually quiet. Maybe they're trying to avoid drawing attention to themselves.
Player3: I see your point about Player4, but it's still early in the game. We should be careful not to jump to conclusions without more evidence.
Player4: I understand your concerns, but I assure you, I'm just a villager. I think we should focus on gathering more information before we start pointing fingers. If the seer has any insights, now might be a good time to share.
Player5: I understand everyone's concerns about Player4, but I think we should also keep an eye on Player1. They were quick to cast suspicion on others, which could be a strategy to divert attention from themselves.
Player6: I understand the concerns about Player4, but I also think we should be cautious about Player1 and Player2. They were quick to agree with each other, which could be a strategy to divert attention from themselves. Let's gather more information before we make any decisions.
Moderator: Now the alive players are Player1, Player2, Player3, Player4, Player5 and Player6. Given the game rules and your role, based on the situation and the information you gain, to vote a player among alive players and to win the game, it's time to vote one player among the alive players, please cast your vote on who you believe is a werewolf. Response in the following format which can be loaded by python json.loads()
{
    "thought": "thought",
    "speak": "player_name"
}
Player1: Player4
Player2: Player4
Player3: Player1
Player4: Player1
Player5: Player1
Player6: Player1
Moderator: Player1 has been voted out.
Moderator: The game goes on.
Moderator: Player2, you are werewolves. If you are alone, eliminate a player, else discuss with your teammates and reach an agreement. Response in the following format which can be loaded by python json.loads(){
    "thought": "thought",
    "speak": "thoughts summary to say to others",
    "agreement": "whether the discussion reached an agreement or not(true/false)"
}
Player2: I think we should consider Player4. They are a strong player and could be a threat.
Moderator: Which player do you vote to kill? Response in the following format which can be loaded by python json.loads()
{{
   "thought": "thought" ,
   "speak": "player_name"
}}
Player2: Player4
Moderator: The player with the most votes is Player4.
Moderator: Would you like to eliminate one player? Response in the following json format which can be loaded by python json.loads()
{{
    "thought": "thought", 
    "speak": "False or player_name"
}}
Player6: False
Moderator: Player5, you're seer. Which player in Player2, Player3, Player4, Player5 and Player6 would you like to check tonight? Response in the following json format which can be loaded by python json.loads()
{
    "thought": "thought" ,
    "speak": "player_name"
}
Player5: Player2
Moderator: Okay, the role of Player2 is werewolf.
Moderator: The day is coming, all the players open your eyes. Last night, the following player(s) has been eliminated: Player4.
Moderator: Now the alive players are Player2, Player3, Player5 and Player6. Given the game rules and your role, based on the situation and the information you gain, to vote a player eliminated among alive players and to win the game, what do you want to say to others? You can decide whether to reveal your role. Response in the following JSON format which can be loaded by python json.loads()
{
    "thought": "thought" ,
    "speak": "thought summary to say to others"
}
Player2: I think we should keep an eye on Player5. They've been pretty quiet, which could mean they're trying to avoid suspicion.
Player3: I think Player2 might be a werewolf. They were quick to suspect Player4, who was eliminated last night. However, the game didn't end after Player1, who also suspected Player4, was voted out. This suggests that Player1 was not a werewolf, and therefore, Player2 is likely to be one.
Player5: I agree with Player3. Player2 was quick to cast suspicion on Player4, who turned out to be innocent. I also noticed that Player2 and Player1 were often in agreement, which could mean they were both werewolves trying to protect each other. I think we should vote out Player2.
Player6: I agree with Player3 and Player5. Player2 was quick to cast suspicion on Player4, who turned out to be innocent. I also noticed that Player2 and Player1 were often in agreement, which could mean they were both werewolves trying to protect each other. I think we should vote out Player2.
Moderator: Now the alive players are Player2, Player3, Player5 and Player6. Given the game rules and your role, based on the situation and the information you gain, to vote a player among alive players and to win the game, it's time to vote one player among the alive players, please cast your vote on who you believe is a werewolf. Response in the following format which can be loaded by python json.loads()
{
    "thought": "thought",
    "speak": "player_name"
}
Player2: Player5
Player3: Player2
Player5: Player2
Player6: Player2
Moderator: Player2 has been voted out.
Moderator: The game is over. The werewolves have been defeated, and the village is safe once again!
\end{lstlisting}

\end{document}